\documentclass[a4paper,aps,prd,10pt,preprintnumbers,showpacs,twocolumn,superscriptaddress,nofootinbib,nobibnotes,floatfix,amsmath,amssymb]{revtex4-1}
\usepackage{graphicx}
\usepackage{cmap}
\usepackage[utf8]{inputenc}
\usepackage[T1]{fontenc}

\begin{document}
\title{Long-lived quasinormal modes of Asymptotically de Sitter Black Holes in Generalized Proca Theory}
\author{S. V. Bolokhov}
\email{bolokhov-sv@rudn.ru}
\affiliation{RUDN University, 6 Miklukho-Maklaya Street, Moscow 117198, Russian Federation}

\begin{abstract}
Massive scalar perturbations of asymptotically de Sitter black holes in generalized Proca theory display a sharp interplay between primary hair, horizon structure, and field mass. Using high-order WKB calculations supplemented by time-domain evolution, we analyze representative black-hole backgrounds and compare the full black-hole spectrum with the exact pure de Sitter benchmark. We show that increasing the scalar mass drives the frequencies into a simple large-mass regime in which the real part grows linearly while the damping rate approaches a nonzero geometry-dependent constant, so true quasi-resonances do not occur within the regime studied here. We also identify how the spectrum shifts with black-hole size and Proca hair, derive a compact analytic large-$\mu$ formula, and comment on the implications of the de Sitter-like sector for strong cosmic censorship in the charged three-horizon regime.
\end{abstract}

\maketitle

\section{Introduction}

Quasinormal modes of black holes  \cite{Kokkotas:1999bd,Berti:2009kk,Konoplya:2011qq,Bolokhov:2025rng} provide a direct probe of the geometry through oscillation frequencies and damping times. For asymptotically de Sitter spacetimes the problem is especially rich because wave propagation is controlled not only by the event horizon but also by the cosmological horizon and by the effective curvature scale of the background.

The asymptotically de Sitter case is especially interesting because the cosmological horizon changes the late-time behavior in a qualitative way: instead of the power-law tails familiar from asymptotically flat geometries, one often finds exponential decay governed by the least-damped quasinormal frequencies, so the quasinormal spectrum is tied more directly to the full time evolution of the perturbation~\cite{Dyatlov:2010hq,Dyatlov:2011asymptotic}. This makes de Sitter black holes a particularly clean setting for identifying the dominant mode, resolving several overtones, and understanding how the cosmological horizon reshapes the spectrum. Quasinormal modes of asymptotically de Sitter black holes have been studied in a number of publications~\cite{Zhidenko:2003wq,Kanti:2005xa,Konoplya:2007jv,Konoplya:2004uk,Cuyubamba:2016cug,Dyatlov:2010hq,Jansen:2017oag,Molina:2003ff,Jing:2003wq,Konoplya:2017ymp,Aragon:2020qdc,Konoplya:2013sba,Mo:2018nnu,Konoplya:2017lhs,Konoplya:2007zx}.

A second reason for the special interest in asymptotically de Sitter quasinormal modes is their connection with strong cosmic censorship. For black holes with an inner Cauchy horizon, the competition between exterior decay and blueshift amplification at that horizon is controlled by the spectral gap of the quasinormal spectrum. One commonly introduces $\alpha_{\rm gap}\equiv \min(-\mathrm{Im}\,\omega)$ and the dimensionless ratio $\beta_{\rm SCC}\equiv \alpha_{\rm gap}/\kappa_-$, where $\kappa_-$ is the surface gravity of the Cauchy horizon. In the smooth-data formulation, Christodoulou-type strong cosmic censorship can be endangered when $\beta_{\rm SCC}>1/2$, while the rough-data formulation leads to the stronger threshold $\beta_{\rm SCC}<1$ as the relevant bound~\cite{Cardoso:2017soq,Dias:2018etb,KonoplyaZhidenko:2022SCC}. Therefore, in the de Sitter problem the quasinormal spectrum is interesting not only as an observable of the exterior geometry but also because it constrains the regularity of possible extensions beyond the Cauchy horizon.

The gravity theory considered here belongs to the generalized Proca class, namely a vector-tensor theory with derivative self-interactions arranged so that the vector field carries only three propagating physical degrees of freedom~\cite{Heisenberg:2014rta}. The particular branch relevant for this work emerges from the recent Proca version of four-dimensional regularized Gauss--Bonnet gravity and admits black-hole solutions with primary hair~\cite{Charmousis:2025jpx,RefProcaGB2026}. In this setup, an integration constant associated with the Proca sector acts as an effective cosmological constant, so asymptotically de Sitter black holes arise even in the absence of a bare cosmological term. This makes the model especially attractive for quasinormal-mode studies, because the same Proca sector that provides independent black-hole hair also determines the asymptotic curvature scale, allowing one to trace directly how primary hair and the effective cosmological constant imprint themselves on the spectrum.

Massive fields deserve separate attention because they can qualitatively modify both the quasinormal spectrum and the late-time evolution. A characteristic effect is the appearance of quasi-resonances, namely modes with extremely small damping rates for which the imaginary part of the frequency tends to zero while the real part remains finite. This phenomenon was first identified for massive scalar perturbations and later extended to fields of different spin, which suggests that it is not tied to a specific test field but reflects a more general property of wave propagation in black-hole backgrounds~\cite{Konoplya:2017tvu,Ohashi:2004wr,Percival:2020skc,Konoplya:2004wg,Fernandes:2021qvr}. Subsequent studies showed that such long-lived modes may arise in a wide class of black-hole spacetimes and even in alternative compact configurations, such as wormholes, provided that the effective potential supports the necessary trapping behavior~\cite{Skvortsova:2026unq,Bolokhov:2024bke,Lutfuoglu:2026xlo,Churilova:2020bql,Zhidenko:2006rs,Lutfuoglu:2025qkt,Bolokhov:2023bwm,Lutfuoglu:2026uzy,Bolokhov:2023ruj,Skvortsova:2025cah,Lutfuoglu:2025hwh,Zinhailo:2018ska,Dubinsky:2025bvf,Bolokhov:2026dzn,Lutfuoglu:2026fpx,Lutfuoglu:2025bsf,Lutfuoglu:2026gis,Skvortsova:2024eqi}. At the same time, quasi-resonances are not universal: in some backgrounds increasing the field mass does not force the damping rate to vanish, showing that the effect depends sensitively on the detailed structure of the potential barrier and on the asymptotic behavior of spacetime~\cite{Zinhailo:2024jzt,Konoplya:2006gq}.

A nonzero field mass also leaves a distinct imprint on the time-domain signal. After the quasinormal-ringing stage, the perturbation may develop an oscillatory late-time tail rather than the familiar power-law decay of massless fields, with the tail frequency and damping governed jointly by the field mass and the background geometry~\cite{Koyama:2001ee,Jing:2004zb,Gibbons:2008rs,Moderski:2001tk,Rogatko:2007zz,Koyama:2000hj,Koyama:2001qw,Gibbons:2008gg,Dubinsky:2024jqi}. For this reason, a consistent treatment of massive perturbations should relate the shape of the effective potential, the quasinormal spectrum, and the transition from ringdown to asymptotic tails within the same framework. In the generalized Proca backgrounds considered here we will show that, although the damping rate decreases strongly as $\mu$ grows, it approaches a nonzero constant rather than zero; hence true quasi-resonances, i.e. arbitrarily long-lived modes, are absent in the large-mass regime.

Motivated by these considerations, in the present work we study a massive test scalar field propagating on the full asymptotically de Sitter black-hole solution of generalized Proca theory~\cite{RefProcaGB2026}. Our aim is to move beyond the pure de Sitter vacuum \cite{Malik:2026oxx} and retain the full black-hole geometry, so as to determine how the field mass, the black-hole parameters, and the Proca-induced effective cosmological scale shape the effective potential, the quasinormal spectrum, and the late-time decay, with particular attention to the large-mass regime where one may ask whether weakly damped modes develop into genuine quasi-resonances. The quasinormal modes of massless fields in this background have been recently analyzed in \cite{Skvortsova:2026idf}. 

The paper is organized as follows. In Sec.~II we review the generalized Proca black-hole background, its horizon structure, and the effective de Sitter asymptotics. In Sec.~III we derive the massive-scalar master equation and analyze the corresponding effective potential. Section~IV summarizes the WKB and time-domain methods used to extract the quasinormal spectrum. The main spectral results are presented in Sec.~V, where we compare the full black-hole frequencies with the exact de Sitter case, discuss the field-mass, charge, and overtone dependences, derive the large-mass asymptotics, and comment on the strong-cosmic-censorship bound. Finally, Sec.~VI summarizes our conclusions and outlines directions for further work.

\section{Generalized Proca Black-Hole Background}

We consider the four-dimensional generalized Proca action relevant for the present primary-hair branch~\cite{Heisenberg:2014rta,Charmousis:2025jpx}
\begin{equation}
\begin{split}
S[g,W]=\int \mathrm{d}^4x\,\sqrt{-g}\Big[
R+c_1 G_{\mu\nu}W^\mu W^\nu \\
+c_2(W^2)^2+c_3 W^2\nabla_\mu W^\mu
\Big],
\end{split}
\end{equation}
where $W_\mu$ is the Proca vector and $W^2\equiv W_\mu W^\mu$.

The Einstein--Hilbert term governs the usual spin-2 geometry, whereas the remaining operators encode nonminimal couplings and derivative self-interactions of the Proca field. Because the vector is massive, there is no Maxwell-type gauge symmetry that would remove its longitudinal mode. This is why, in the static and spherically symmetric sector considered here, the vector field can consistently carry both temporal and radial components.

For static spherical symmetry we write
\begin{equation}
\begin{split}
\mathrm{d}s^2&=-f(r)\,\mathrm{d}t^2+\frac{\mathrm{d}r^2}{f(r)}+r^2\mathrm{d}\Omega_2^2,\\
W_\mu \mathrm{d}x^\mu&=w_0(r)\,\mathrm{d}t+w_1(r)\,\mathrm{d}r,
\end{split}
\end{equation}
Here $w_0(r)$ is the temporal, electric-type component of the Proca field, whereas $w_1(r)$ is its radial longitudinal component. The branch condition
\begin{equation}
W^2=\lambda=\text{constant}
\end{equation}
selects the exact family of hairy solutions for which the vector norm is frozen to a constant and therefore introduces a preferred background scale into the geometry~\cite{Heisenberg:2014rta,Charmousis:2025jpx,RefProcaGB2026}.
It is convenient to define the composite parameters
\begin{equation}
\alpha\equiv -\frac{c_1^3}{c_3^2},
\qquad
\beta\equiv \left(1-\frac{8c_1c_2}{3c_3^2}\right)c_1,
\end{equation}
\begin{equation}
A\equiv 1-\frac{\beta\lambda}{2},
\qquad
B\equiv 1-\beta\lambda\left(1-\frac{c_1\lambda}{4}\right).
\end{equation}

The full metric function of the asymptotically de Sitter case  derived in Ref.~\cite{RefProcaGB2026} can then be written as
\begin{equation}
\label{eq:fullmetric}
\begin{split}
f(r)&=\,1-\frac{2(M-Q)}{r}\\
&+\frac{r^2}{2\alpha}\Bigg(A
-\sqrt{B+\frac{8\alpha}{r^3}\left[Q+\frac{\lambda(M-Q)}{2\beta}\right]}\Bigg).
\end{split}
\end{equation}
The parameters $(M,Q)$ control the black-hole sector, while the combinations $(\alpha,\beta,c_1,\lambda)$ control the effective de Sitter asymptotics.

It is useful to separate the physical meaning of these constants. The parameter $M$ is the mass-type integration constant of the solution: increasing $M$ strengthens the attractive part of the geometry and typically enlarges the black-hole horizon. By contrast, $Q$ is not a Maxwell charge. Rather, it is an independent primary-hair parameter associated with the Proca sector \cite{Charmousis:2025jpx,RefProcaGB2026}. The couplings $(\alpha,\beta,c_1)$ together with the constant norm $\lambda$ determine how strongly this vector hair backreacts on the metric and how the effective cosmological scale is generated.

The large-$r$ expansion of Eq.~\eqref{eq:fullmetric} makes this interpretation more transparent. Writing
\begin{equation}
f(r)=1-H^2r^2-\frac{2\mathcal{M}}{r}+\mathcal{O}(r^{-4}),
\label{eq:asymptoticf}
\end{equation}
one finds
\begin{equation}
\mathcal{M}\equiv (M-Q)+\frac{1}{\sqrt{B}}\left[Q+\frac{\lambda(M-Q)}{2\beta}\right].
\label{eq:masscoeff}
\end{equation}
Thus $H^{-1}$ is the effective de Sitter curvature radius, while the coefficient of the Schwarzschild-like $1/r$ tail is controlled by a definite combination of $M$ and $Q$. In this sense $M$ sets the overall mass scale, whereas $Q$ measures an additional deformation due to Proca hair. The pure de Sitter limit is recovered by setting $M=Q=0$, for which $f(r)=1-H^2r^2$.

In terms of these quantities, the effective cosmological constant is encoded through
\begin{equation}
\Lambda_{\rm eff}=3H^2=\frac{3(\sqrt{B}-A)}{2\alpha},
\end{equation}
whenever the parameters are such that the geometry possesses an event horizon $r_h$ and a cosmological horizon $r_c>r_h$.

The horizon structure is determined by the roots of the metric function,
\begin{equation}
f(r_i)=0,
\qquad
r_i\in\{r_-,r_h,r_c\},
\end{equation}
where $r_h$ denotes the event horizon, $r_c$ the cosmological horizon, and $r_-$ an inner Cauchy horizon whenever such a root is present. The physically relevant static patch for the perturbation problem is therefore the finite interval
\begin{equation}
r_h<r<r_c,
\qquad
f(r)>0,
\end{equation}
in which the Killing vector $\partial_t$ remains timelike. The associated surface gravities,
\begin{equation}
\kappa_i=\frac{1}{2}\left|f'(r_i)\right|,
\end{equation}
set the redshift scales at the boundaries. They enter the tortoise coordinate, the quasinormal-mode boundary conditions, and later the strong-cosmic-censorship measure based on the ratio $\alpha_{\rm gap}/\kappa_-$.

\section{Massive Scalar Perturbations}

We study a minimally coupled scalar field of mass $\mu$ satisfying
\begin{equation}
\left(\Box-\mu^2\right)\Phi=0.
\end{equation}
Using the standard decomposition \cite{Carter:1968ks,Konoplya:2018arm} 
\begin{equation}
\Phi=e^{-i\omega t}Y_{\ell m}(\theta,\varphi)\frac{U(r)}{r},
\end{equation}
we obtain the radial equation
\begin{equation}
\frac{1}{r^2}\frac{\mathrm{d}}{\mathrm{d}r}\left(r^2f(r)\frac{\mathrm{d}}{\mathrm{d}r}\frac{U}{r}\right)+\left(\frac{\omega^2}{f(r)}-\frac{\ell(\ell+1)}{r^2}-\mu^2\right)\frac{U}{r}=0.
\end{equation}

Introducing the tortoise coordinate,
\begin{equation}
\frac{\mathrm{d}r_*}{\mathrm{d}r}=\frac{1}{f(r)},
\end{equation}
one finds the Schr\"odinger-type master equation
\begin{equation}
\frac{\mathrm{d}^2U}{\mathrm{d}r_*^2}+\left[\omega^2-V(r)\right]U=0,
\end{equation}
with effective potential
\begin{equation}
\label{eq:potential}
V(r)=f(r)\left[\frac{\ell(\ell+1)}{r^2}+\frac{f'(r)}{r}+\mu^2\right].
\end{equation}
This form makes transparent how the field mass, centrifugal barrier, and black-hole geometry all enter the perturbation problem.

To make the role of the effective cosmological term explicit, it is useful to compare two simple representative examples within the same generalized Proca family. In both examples we set
\begin{equation}
(\alpha,\beta,\lambda,Q,M,\ell)=(1,1,0.2,0,1,1),
\end{equation}
and vary only $c_1$. These values are chosen purely for illustration: they provide simple representative backgrounds for comparing the Schwarzschild-like and asymptotically de Sitter sectors, without singling out a special point in parameter space.

\begin{figure*}[t]
\centering
\includegraphics[width=0.98\textwidth]{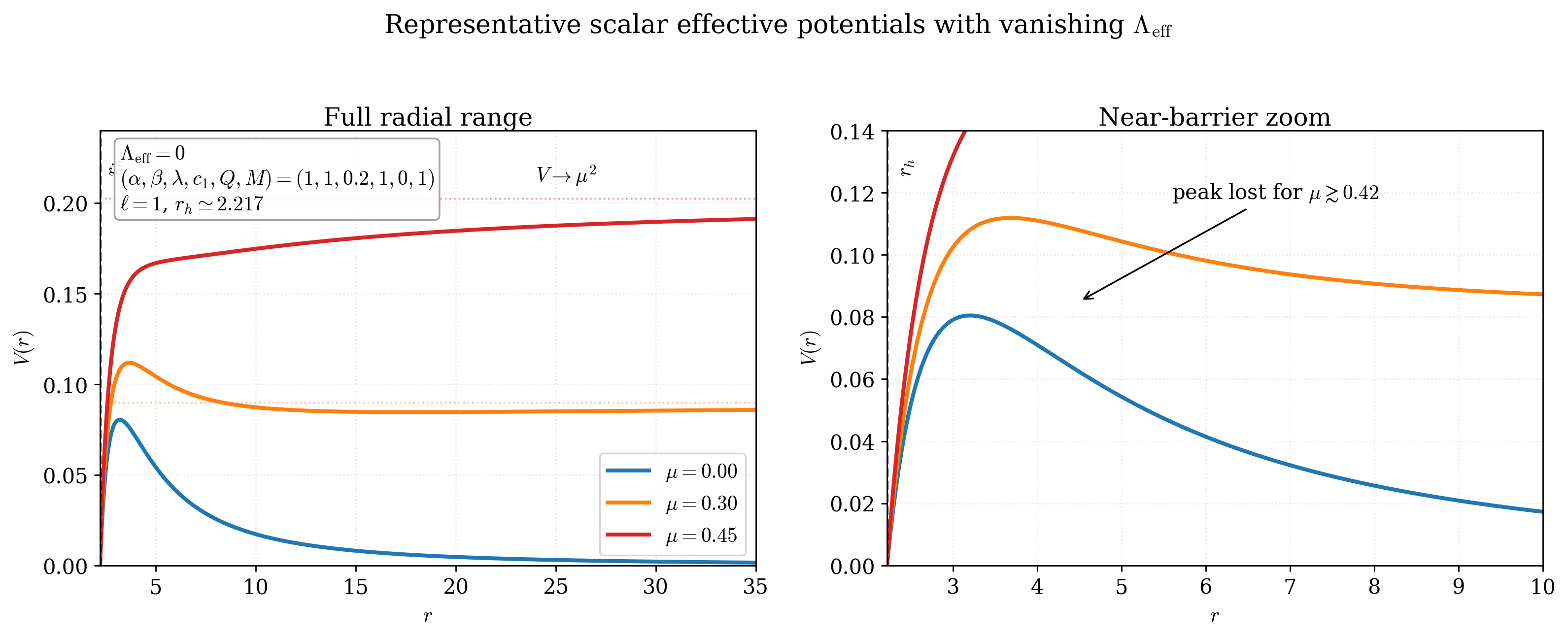}
\caption{Representative effective potentials for the $\Lambda_{\rm eff}=0$ benchmark $(\alpha,\beta,\lambda,c_1,Q,M,\ell)=(1,1,0.2,1,0,1,1)$. Left: full radial range outside the event horizon $r_h\simeq 2.217$. Right: zoom near the barrier. As $\mu$ increases, the asymptotic plateau $V\to\mu^2$ is lifted, and the local peak is lost for $\mu\gtrsim 0.42$ in this example.}
\label{fig:veff-flat}
\end{figure*}

For $c_1=1$ one has $\Lambda_{\rm eff}=0$ and a single event horizon at $r_h\simeq 2.217$. In this case the potential approaches a mass plateau,
\begin{equation}
V(r)\longrightarrow \mu^2,\qquad r\to\infty,
\end{equation}
so increasing the scalar mass raises the asymptotic tail. Figure~\ref{fig:veff-flat} shows that for this representative $\ell=1$ case the barrier peak disappears once the field mass becomes sufficiently large, numerically around $\mu\approx 0.42$.

\begin{figure*}[t]
\centering
\includegraphics[width=0.86\textwidth]{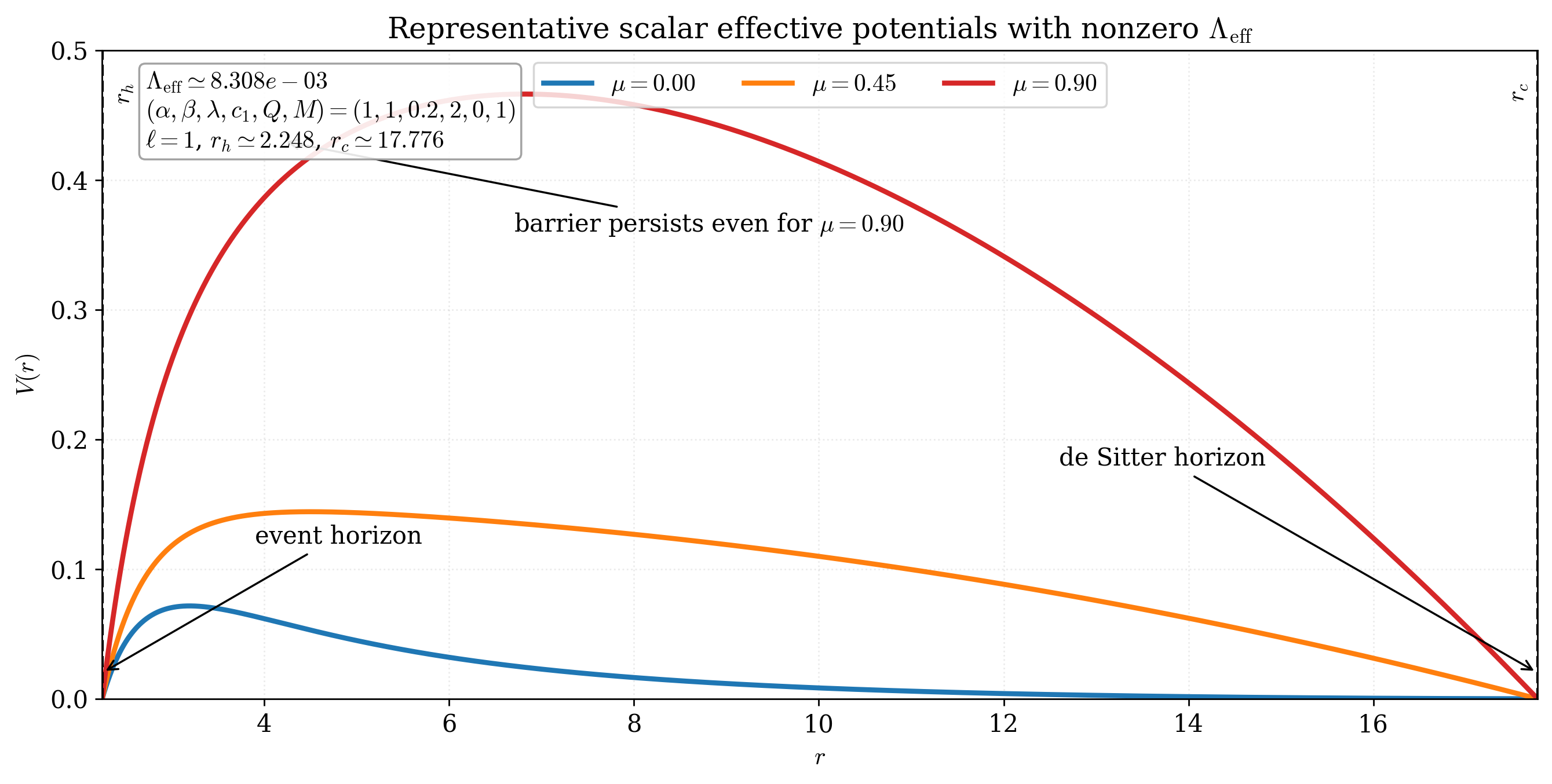}
\caption{Representative effective potentials for the de Sitter benchmark $(\alpha,\beta,\lambda,c_1,Q,M,\ell)=(1,1,0.2,2,0,1,1)$, for which $\Lambda_{\rm eff}\simeq 8.31\times10^{-3}$. The event horizon $r_h\simeq 2.248$ and the de Sitter horizon $r_c\simeq 17.776$ are indicated explicitly. Unlike the $\Lambda_{\rm eff}=0$ case, the potential vanishes at both horizons and keeps a well-defined peak between them even for comparatively large scalar masses.}
\label{fig:veff-ds}
\end{figure*}

For $c_1=2$ one instead obtains an asymptotically de Sitter configuration with
\begin{equation}
\Lambda_{\rm eff}\simeq 8.31\times 10^{-3},
\end{equation}
and a static region bounded by an event horizon and a de Sitter horizon,
\begin{equation}
r_h\simeq 2.248,
\qquad
r_c\simeq 17.776.
\end{equation}
Because $f(r)$ vanishes at both horizons, the effective potential also decays to zero at both ends of the interval $(r_h,r_c)$. As illustrated in Fig.~\ref{fig:veff-ds}, the potential therefore retains a barrier peak across the static patch even for scalar masses that are already above the flat-space threshold.

\section{Methods for Finding Quasinormal Modes}

For the asymptotically de Sitter black-hole geometry, quasinormal modes are defined by purely ingoing behavior at the event horizon and purely outgoing behavior at the cosmological horizon:
\begin{equation}
U\sim e^{-i\omega r_*}, \qquad r\to r_h,
\end{equation}
\begin{equation}
U\sim e^{+i\omega r_*}, \qquad r\to r_c.
\end{equation}
Since $r_*\to -\infty$ at $r=r_h$ and $r_*\to +\infty$ at $r=r_c$, these boundary conditions select a discrete set of complex frequencies $\omega$. For the present problem, two complementary methods are especially natural: a WKB treatment of the barrier region and a time-domain characteristic evolution following Gundlach et al.~\cite{GundlachPricePullin:1994,GundlachPricePullin:1994b}.

\subsection{WKB method}

When the effective potential is smooth and has a single dominant peak between the two horizons, the master equation can be treated semiclassically. Let $r_*^{(0)}$ denote the location of the potential maximum, and define
\begin{equation}
V_0\equiv V\bigl(r_*^{(0)}\bigr),
\qquad
V_0^{(m)}\equiv \left.\frac{\mathrm{d}^mV}{\mathrm{d}r_*^m}\right|_{r_*=r_*^{(0)}}.
\end{equation}
At $N$th WKB order the quasinormal frequencies satisfy the standard quantization condition~\cite{Konoplya:2011qq,Matyjasek:2017psv,Matyjasek:2026yiu}
\begin{equation}
\frac{i\bigl(\omega^2-V_0\bigr)}{\sqrt{-2V_0''}}-\sum_{j=2}^{N}\Lambda_j=n+\frac{1}{2},
\qquad n=0,1,2,\ldots,
\end{equation}
where the correction terms $\Lambda_j$ are known functions of the higher derivatives $V_0^{(m)}$. At leading order this reduces to
\begin{equation}
\omega^2\approx V_0-i\left(n+\frac{1}{2}\right)\sqrt{-2V_0''}.
\end{equation}
This method is particularly useful for obtaining the fundamental and low-lying overtones of single-peaked barriers, especially for moderate and large angular momentum. Consequently, this method has been widely employed for the computation of quasinormal modes \cite{Konoplya:2009hv,Malik:2026lfj,Bolokhov:2025egl,Eniceicu:2019npi,Fernando:2016ftj,Guo:2020caw,Pathrikar:2025gzu,Kokkotas:2010zd,Gonzalez:2022ote,Bolokhov:2025lnt,Bolokhov:2025aqy,Momennia:2018hsm,Malik:2025erb}, and has also been applied to the evaluation of grey-body factors for black holes in a number of recent works \cite{Dubinsky:2025nxv,Lutfuoglu:2025ohb,Malik:2025czt,Skvortsova:2024msa,Konoplya:2023ppx,Dubinsky:2026wcv,Lutfuoglu:2025ljm,Dubinsky:2024nzo,Lutfuoglu:2025hjy,Konoplya:2019ppy,Dubinsky:2025wns,Malik:2025qnr,Lutfuoglu:2025blw,Dubinsky:2025ypj,Konoplya:2023moy}. In the present generalized Proca problem, the WKB scheme is therefore most reliable for asymptotically de Sitter black-hole backgrounds, where the potential vanishes at both horizons and keeps a barrier in the static patch. By contrast, when $\Lambda_{\rm eff}=0$ and the massive-field plateau removes the peak, the WKB turning-point picture ceases to be reliable, so the method should then be used with caution or abandoned in favor of direct time evolution.

\subsection{Time-domain integration following Gundlach et al.}

A more robust approach is to evolve the perturbation directly in the time domain. Introducing the null coordinates
\begin{equation}
u=t-r_*,
\qquad
v=t+r_*,
\end{equation}
the master equation takes the characteristic form
\begin{equation}
4\frac{\partial^2U}{\partial u\,\partial v}+V\bigl(r(u,v)\bigr)U=0.
\end{equation}
One then prescribes initial data on two intersecting null segments, for instance a Gaussian pulse on one segment and trivial data on the other, and evolves the field on a null grid.

For a null rectangle with South, East, West, and North points denoted by $S$, $E$, $W$, and $N$, the Gundlach--Price--Pullin integration scheme is
\begin{equation}
U_N=U_W+U_E-U_S-\frac{\Delta^2}{8}V_S\bigl(U_W+U_E\bigr)+\mathcal{O}(\Delta^4),
\end{equation}
where $\Delta$ is the null-grid spacing and $V_S$ is the potential evaluated at the South point. Repeating this update across the grid produces the full waveform in the static region between the horizons. The quasinormal ringing is then read off from the intermediate-time signal at fixed radius, or along one of the horizons, by fitting the data to a superposition
\begin{equation}
U(t)\sim \sum_k A_k e^{-i\omega_k t},
\end{equation}
from which the dominant complex frequencies can be extracted in practice by Prony-type fitting or related signal-processing techniques~\cite{Dubinsky:2024gwo}.

The time-domain method has two important advantages for the present problem. First, it does not assume in advance that the potential has the ideal single-barrier structure required by the WKB approximation, and it has therefore been successfully applied to a wide variety of black-hole and wormhole configurations~\cite{Abdalla:2012si,Malik:2023bxc,Qian:2022kaq,Arbelaez:2025gwj,Konoplya:2005et,Dubinsky:2025fwv,Skvortsova:2024atk,Konoplya:2024kih,Bolokhov:2024ixe,Arbelaez:2026eaz,Aneesh:2018hlp,Konoplya:2018yrp,Skvortsova:2023zmj,Malik:2024nhy,Bolokhov:2023dxq,Konoplya:2014lha,Varghese:2011ku,Konoplya:2024hfg,Skvortsova:2024wly,Konoplya:2023aph,Bolokhov:2023bwm}. Second, it directly reveals whether the perturbation decays, oscillates, or develops a late-time tail, and is therefore well suited for stability studies in parameter regions where the effective potential is strongly deformed by the Proca couplings or by the scalar-field mass.

\section{Quasinormal modes}

Keeping the couplings $(\alpha,\beta,c_1,\lambda)$ fixed and switching off the black-hole data $(M,Q)$ in Eq.~(\ref{eq:fullmetric}) yields the pure de Sitter vacuum
\begin{equation}
f_{\rm dS}(r)=1-H^2r^2,
\qquad
H^2=\frac{\Lambda_{\rm eff}}{3}=\frac{\sqrt{B}-A}{2\alpha},
\end{equation}
with a single cosmological horizon at $r_c^{\rm(dS)}=H^{-1}$. In this limit the massive Klein--Gordon problem is exactly solvable, and the scalar quasinormal spectrum is~\cite{Malik:2026aqs}
\begin{equation}
\label{eq:ds_exact_qnm}
\omega_{n\ell}^{(\pm)}=-iH\left(2n+\ell+\frac{3}{2}\pm\sqrt{\frac{9}{4}-\frac{\mu^2}{H^2}}\right).
\end{equation}
for light fields with $\mu\le 3H/2$ and overtone number $n=0,1,2,\ldots$. The massless limit gives two purely damped branches,
\begin{equation}
\label{eq:ds_massless_qnm}
\omega_{n\ell}^{(1)}=-iH(2n+\ell),
\qquad
\omega_{n\ell}^{(2)}=-iH(2n+\ell+3).
\end{equation}
For heavier fields, $\mu>3H/2$, the square root becomes imaginary and the exact de Sitter modes acquire an oscillatory part,
\begin{equation}
\label{eq:ds_heavy_qnm}
\omega_{n\ell}^{(\pm)}=\pm\sqrt{\mu^2-\frac{9H^2}{4}}-iH\left(2n+\ell+\frac{3}{2}\right).
\end{equation}

These formulas provide a natural analytic benchmark for the numerical treatment of the full black-hole geometry. In particular, for the same couplings used in the de Sitter potential example of Fig.~\ref{fig:veff-ds}, namely $(\alpha,\beta,\lambda,c_1)=(1,1,0.2,2)$, one finds
\begin{equation}
\begin{aligned}
H&\simeq 5.26\times 10^{-2},\\
r_c^{\rm(dS)}&=H^{-1}\simeq 19.00,\\
\frac{3H}{2}&\simeq 7.89\times 10^{-2}.
\end{aligned}
\end{equation}
The exact vacuum therefore predicts, for example, the massless fundamental $\ell=1$ frequencies
\begin{equation}
\omega_{0,1}^{(1)}\simeq -5.26\times 10^{-2}i,
\qquad
\omega_{0,1}^{(2)}\simeq -2.10\times 10^{-1}i,
\end{equation}
while for a heavier field such as $\mu=0.10$ one obtains
\begin{equation}
\omega_{0,1}^{(\pm)}\simeq \pm 6.14\times 10^{-2}-1.32\times 10^{-1}i.
\end{equation}
These values give a clean analytic target for the de Sitter-like sector of the problem and help distinguish it from the Schwarzschild-like sector in the full black-hole geometry.

A useful way to organize the full asymptotically de Sitter spectrum is to separate it into two branches~\cite{KonoplyaZhidenko:2022Nonosc}. One branch is Schwarzschild-like: it is perturbative in $\Lambda_{\rm eff}$ and continuously reduces to the ordinary asymptotically flat black-hole quasinormal spectrum when $\Lambda_{\rm eff}\to 0$. The other branch is de Sitter-like: it is continuously connected to the empty de Sitter frequencies in Eqs.~\eqref{eq:ds_exact_qnm}--\eqref{eq:ds_heavy_qnm}, but deformed by the presence of the black hole. For clarity, the tables reported below list the Schwarzschild branch of modes. By contrast, the de Sitter branch is expected to show up most clearly in the time domain through the exponential late-time tail, because for asymptotically de Sitter geometries the late-time decay is exponential rather than the power-law tail familiar from asymptotically flat backgrounds~\cite{KonoplyaZhidenko:2022Nonosc,Konoplya:2011qq}.

Once $M$ and $Q$ are restored, the exact spectrum \eqref{eq:ds_exact_qnm} is no longer expected to describe every mode, but it still provides the natural analytic reference for the de Sitter-like branch in the small-black-hole regime. To make that statement operational, it is useful to freeze the asymptotic sector at the same couplings used in Fig.~\ref{fig:veff-ds},
\begin{equation}
(\alpha,\beta,\lambda,c_1)=(1,1,0.2,2),
\qquad
\Lambda_{\rm eff}\simeq 8.31\times 10^{-3},
\end{equation}
and then compute quasinormal modes for a short list of representative black-hole backgrounds:
\begin{itemize}
\item \textit{Small neutral black hole:} $(M,Q)=(0.50,0)$, with
$(r_h,r_c)\simeq (1.106,18.421)$ and $r_h/r_c\simeq 0.060$.
This is the cleanest case for checking how rapidly the dominant mode approaches the exact de Sitter prediction.
\item \textit{Reference neutral black hole:} $(M,Q)=(1.00,0)$, with
$(r_h,r_c)\simeq (2.248,17.776)$ and $r_h/r_c\simeq 0.126$.
This is the natural continuation of the potential benchmark already shown in Fig.~\ref{fig:veff-ds}.
\item \textit{Moderately charged black hole:} $(M,Q)=(1.00,0.60)$, with
$(r_h,r_c)\simeq (2.045,17.781)$ and $r_h/r_c\simeq 0.115$.
This point isolates the effect of charge at fixed cosmological scale and fixed mass.
\item \textit{Strongly charged black hole:} $(M,Q)=(1.00,0.90)$, with
$(r_h,r_c)\simeq (1.736,17.783)$ and $r_h/r_c\simeq 0.098$.
This provides a more stringent charge-dependence test while still keeping a clean event-horizon/cosmological-horizon configuration.
\item \textit{Large neutral black hole:} $(M,Q)=(3.20,0)$, with
$(r_h,r_c)\simeq (9.423,12.448)$ and $r_h/r_c\simeq 0.757$.
This is a near-Nariai-like regime in which the two outer horizons are much closer, so one expects the spectrum to deviate more strongly from the pure de Sitter pattern.
\end{itemize}

These five points give a practical first scan of the full geometry. The first two are best for quantifying how the dominant frequencies depart from the pure de Sitter reference when the black hole is small; the third and fourth track how the spectrum shifts as $Q$ increases at fixed asymptotics; and the fifth tests the opposite regime of a much larger black hole. The de Sitter branch itself is expected to be identified more cleanly from the exponentially decaying late-time tail of the time-domain signal than from the WKB tables. Once those spectra are in hand, the field-mass dependence can be added systematically, for example by repeating each run for a light field ($\mu<3H/2$) and for a heavier field ($\mu>3H/2$).

In this way the comparison with the exact de Sitter benchmark becomes fully concrete: the same Proca couplings determine $H$ and $\Lambda_{\rm eff}$, while $(M,Q)$ move the horizons and reshape the barrier. The resulting quasinormal frequencies therefore separate the Schwarzschild-like sector, which is continuously connected to the asymptotically flat limit, from the de Sitter-like sector, which is continuously connected to the empty de Sitter vacuum.

\begin{table*}[t]
\caption{Fundamental ($n=0$) massive-scalar quasinormal frequencies for the small neutral generalized Proca black-hole benchmark with $(\alpha,\beta,\lambda,c_1,M,Q)=(1,1,0.2,2,0.50,0)$, $\Lambda_{\rm eff}\simeq 8.31\times 10^{-3}$, $r_h\simeq 1.106$, and $r_c\simeq 18.421$. The three angular sectors $\ell=0,1,2$ are displayed side by side for the same set of field masses $\mu$. In each block, $\omega^{(16)}$ denotes the 16th-order WKB value with Pad\'e approximant $\tilde m=8$, $\omega^{(14)}$ denotes the 14th-order value with $\tilde m=7$, and $\Delta$ is the relative difference between the two approximants, in percent.}
\label{tab:small-neutral-bh-wkb}
\centering
\scriptsize
\setlength{\tabcolsep}{3.5pt}
\renewcommand{\arraystretch}{1.05}
\resizebox{\textwidth}{!}{%
\begin{tabular}{c c c c c c c c c c}
\hline\hline
 & \multicolumn{3}{c}{$\ell=0$} & \multicolumn{3}{c}{$\ell=1$} & \multicolumn{3}{c}{$\ell=2$} \\
\cline{2-4}\cline{5-7}\cline{8-10}
$\mu$ & $\omega^{(16)}$ & $\omega^{(14)}$ & $\Delta(\%)$ & $\omega^{(16)}$ & $\omega^{(14)}$ & $\Delta(\%)$ & $\omega^{(16)}$ & $\omega^{(14)}$ & $\Delta(\%)$ \\
\hline
$0$ & $0.196749-0.187630\,i$ & $0.196844-0.187487\,i$ & $0.0632$ & $0.521864-0.173755\,i$ & $0.521864-0.173756\,i$ & $0.00007$ & $0.862510-0.171763\,i$ & $0.862510-0.171763\,i$ & $0$ \\
$0.5$ & $0.423561-0.035806\,i$ & $0.423564-0.035802\,i$ & $0.00123$ & $0.584637-0.135394\,i$ & $0.584637-0.135394\,i$ & $0$ & $0.906141-0.157123\,i$ & $0.906142-0.157123\,i$ & $0.00001$ \\
$1.0$ & $0.846576-0.037853\,i$ & $0.846575-0.037854\,i$ & $0.00008$ & $0.876612-0.029039\,i$ & $0.877077-0.030264\,i$ & $0.149$ & $1.042513-0.108904\,i$ & $1.042521-0.108875\,i$ & $0.00293$ \\
$1.5$ & $1.269279-0.038242\,i$ & $1.269279-0.038242\,i$ & $0$ & $1.287205-0.036431\,i$ & $1.287205-0.036431\,i$ & $0.00002$ & $1.343836-0.034300\,i$ & $1.343836-0.034300\,i$ & $7.4\times 10^{-6}$ \\
$2.0$ & $1.692065-0.038376\,i$ & $1.692065-0.038376\,i$ & $0$ & $1.705000-0.037521\,i$ & $1.705000-0.037521\,i$ & $0$ & $1.734531-0.034840\,i$ & $1.734531-0.034840\,i$ & $0$ \\
$2.5$ & $2.114899-0.038437\,i$ & $2.114899-0.038437\,i$ & $0$ & $2.125070-0.037928\,i$ & $2.125070-0.037928\,i$ & $0$ & $2.147006-0.036637\,i$ & $2.147006-0.036637\,i$ & $0$ \\
$3.0$ & $2.537758-0.038469\,i$ & $2.537758-0.038469\,i$ & $0$ & $2.546157-0.038129\,i$ & $2.546157-0.038129\,i$ & $0$ & $2.563806-0.037336\,i$ & $2.563806-0.037336\,i$ & $0$ \\
$3.5$ & $2.960632-0.038489\,i$ & $2.960632-0.038489\,i$ & $0$ & $2.967792-0.038244\,i$ & $2.967792-0.038244\,i$ & $0$ & $2.982625-0.037699\,i$ & $2.982625-0.037699\,i$ & $0$ \\
$4.0$ & $3.383516-0.038502\,i$ & $3.383516-0.038502\,i$ & $0$ & $3.389759-0.038317\,i$ & $3.389759-0.038317\,i$ & $0$ & $3.402580-0.037916\,i$ & $3.402580-0.037916\,i$ & $0$ \\
$4.5$ & $3.806407-0.038510\,i$ & $3.806407-0.038510\,i$ & $0$ & $3.811943-0.038365\,i$ & $3.811943-0.038365\,i$ & $0$ & $3.823246-0.038058\,i$ & $3.823246-0.038058\,i$ & $0$ \\
$5.0$ & $4.229302-0.038516\,i$ & $4.229302-0.038516\,i$ & $0$ & $4.234276-0.038400\,i$ & $4.234276-0.038400\,i$ & $0$ & $4.244390-0.038155\,i$ & $4.244390-0.038155\,i$ & $0$ \\
$5.5$ & $4.652200-0.038521\,i$ & $4.652200-0.038521\,i$ & $0$ & $4.656717-0.038425\,i$ & $4.656717-0.038425\,i$ & $0$ & $4.665873-0.038226\,i$ & $4.665873-0.038226\,i$ & $0$ \\
$6.0$ & $5.075101-0.038524\,i$ & $5.075101-0.038524\,i$ & $0$ & $5.079237-0.038444\,i$ & $5.079237-0.038444\,i$ & $0$ & $5.087604-0.038278\,i$ & $5.087604-0.038278\,i$ & $0$ \\
$6.5$ & $5.498004-0.038527\,i$ & $5.498004-0.038527\,i$ & $0$ & $5.501819-0.038459\,i$ & $5.501819-0.038459\,i$ & $0$ & $5.509523-0.038319\,i$ & $5.509523-0.038319\,i$ & $0$ \\
$7.0$ & $5.920908-0.038529\,i$ & $5.920908-0.038529\,i$ & $0$ & $5.924449-0.038471\,i$ & $5.924449-0.038471\,i$ & $0$ & $5.931589-0.038350\,i$ & $5.931589-0.038350\,i$ & $0$ \\
$7.5$ & $6.343814-0.038531\,i$ & $6.343814-0.038531\,i$ & $0$ & $6.347117-0.038480\,i$ & $6.347117-0.038480\,i$ & $0$ & $6.353771-0.038376\,i$ & $6.353771-0.038376\,i$ & $0$ \\
$8.0$ & $6.766720-0.038532\,i$ & $6.766720-0.038532\,i$ & $0$ & $6.769816-0.038488\,i$ & $6.769816-0.038488\,i$ & $0$ & $6.776046-0.038396\,i$ & $6.776046-0.038396\,i$ & $0$ \\
$8.5$ & $7.189628-0.038534\,i$ & $7.189628-0.038534\,i$ & $0$ & $7.192540-0.038494\,i$ & $7.192540-0.038494\,i$ & $0$ & $7.198398-0.038413\,i$ & $7.198398-0.038413\,i$ & $0$ \\
$9.0$ & $7.612536-0.038535\,i$ & $7.612536-0.038535\,i$ & $0$ & $7.615286-0.038499\,i$ & $7.615286-0.038499\,i$ & $0$ & $7.620813-0.038428\,i$ & $7.620813-0.038428\,i$ & $0$ \\
$9.5$ & $8.035444-0.038535\,i$ & $8.035444-0.038535\,i$ & $0$ & $8.038049-0.038504\,i$ & $8.038049-0.038504\,i$ & $0$ & $8.043281-0.038440\,i$ & $8.043281-0.038440\,i$ & $0$ \\
$10.0$ & $8.458353-0.038536\,i$ & $8.458353-0.038536\,i$ & $0$ & $8.460827-0.038508\,i$ & $8.460827-0.038508\,i$ & $0$ & $8.465795-0.038450\,i$ & $8.465795-0.038450\,i$ & $0$ \\
\hline\hline
\end{tabular}%
}
\end{table*}

\begin{table*}[t]
\caption{Fundamental ($n=0$) massive-scalar quasinormal frequencies for the reference neutral generalized Proca black-hole benchmark with $(\alpha,\beta,\lambda,c_1,M,Q)=(1,1,0.2,2,1.00,0)$, $\Lambda_{\rm eff}\simeq 8.31\times 10^{-3}$, $r_h\simeq 2.248$, and $r_c\simeq 17.776$. The three angular sectors $\ell=0,1,2$ are displayed side by side for the same set of field masses $\mu$. In each block, $\omega^{(16)}$ denotes the 16th-order WKB value with Pad\'e approximant $\tilde m=8$, $\omega^{(14)}$ denotes the 14th-order value with $\tilde m=7$, and $\Delta$ is the relative difference between the two approximants, in percent.}
\label{tab:reference-neutral-bh-wkb}
\centering
\scriptsize
\setlength{\tabcolsep}{3.5pt}
\renewcommand{\arraystretch}{1.05}
\resizebox{\textwidth}{!}{%
\begin{tabular}{c c c c c c c c c c}
\hline\hline
 & \multicolumn{3}{c}{$\ell=0$} & \multicolumn{3}{c}{$\ell=1$} & \multicolumn{3}{c}{$\ell=2$} \\
\cline{2-4}\cline{5-7}\cline{8-10}
$\mu$ & $\omega^{(16)}$ & $\omega^{(14)}$ & $\Delta(\%)$ & $\omega^{(16)}$ & $\omega^{(14)}$ & $\Delta(\%)$ & $\omega^{(16)}$ & $\omega^{(14)}$ & $\Delta(\%)$ \\
\hline
$0$ & $0.094568-0.094048\,i$ & $0.094607-0.093999\,i$ & $0.0472$ & $0.249251-0.084986\,i$ & $0.249251-0.084986\,i$ & $0$ & $0.413699-0.083425\,i$ & $0.413699-0.083425\,i$ & $0$ \\
$0.5$ & $0.370213-0.032471\,i$ & $0.370061-0.032272\,i$ & $0.0672$ & $0.409030-0.034093\,i$ & $0.409014-0.034132\,i$ & $0.0103$ & $0.503128-0.055840\,i$ & $0.503128-0.055840\,i$ & $0.00005$ \\
$1.0$ & $0.740418-0.033482\,i$ & $0.740418-0.033483\,i$ & $0.00005$ & $0.754984-0.032577\,i$ & $0.754984-0.032575\,i$ & $0.00025$ & $0.789108-0.031469\,i$ & $0.789109-0.031464\,i$ & $0.00058$ \\
$1.5$ & $1.110647-0.033631\,i$ & $1.110647-0.033631\,i$ & $0$ & $1.119981-0.033271\,i$ & $1.119981-0.033272\,i$ & $0$ & $1.139704-0.032521\,i$ & $1.139704-0.032521\,i$ & $0$ \\
$2.0$ & $1.480875-0.033680\,i$ & $1.480875-0.033680\,i$ & $0$ & $1.487789-0.033484\,i$ & $1.487789-0.033484\,i$ & $0$ & $1.502020-0.033081\,i$ & $1.502020-0.033081\,i$ & $0$ \\
$2.5$ & $1.851102-0.033702\,i$ & $1.851102-0.033702\,i$ & $0$ & $1.856603-0.033578\,i$ & $1.856603-0.033578\,i$ & $0$ & $1.867801-0.033326\,i$ & $1.867801-0.033326\,i$ & $0$ \\
$3.0$ & $2.221329-0.033714\,i$ & $2.221329-0.033714\,i$ & $0$ & $2.225899-0.033629\,i$ & $2.225899-0.033629\,i$ & $0$ & $2.235151-0.033456\,i$ & $2.235151-0.033456\,i$ & $0$ \\
$3.5$ & $2.591554-0.033721\,i$ & $2.591554-0.033721\,i$ & $0$ & $2.595464-0.033659\,i$ & $2.595464-0.033659\,i$ & $0$ & $2.603355-0.033533\,i$ & $2.603355-0.033533\,i$ & $0$ \\
$4.0$ & $2.961780-0.033726\,i$ & $2.961780-0.033726\,i$ & $0$ & $2.965197-0.033678\,i$ & $2.965197-0.033678\,i$ & $0$ & $2.972078-0.033582\,i$ & $2.972078-0.033582\,i$ & $0$ \\
$4.5$ & $3.332004-0.033729\,i$ & $3.332004-0.033729\,i$ & $0$ & $3.335040-0.033691\,i$ & $3.335040-0.033691\,i$ & $0$ & $3.341143-0.033616\,i$ & $3.341143-0.033616\,i$ & $0$ \\
$5.0$ & $3.702229-0.033731\,i$ & $3.702229-0.033731\,i$ & $0$ & $3.704959-0.033701\,i$ & $3.704959-0.033701\,i$ & $0$ & $3.710444-0.033640\,i$ & $3.710444-0.033640\,i$ & $0$ \\
$5.5$ & $4.072454-0.033733\,i$ & $4.072454-0.033733\,i$ & $0$ & $4.074935-0.033708\,i$ & $4.074935-0.033708\,i$ & $0$ & $4.079915-0.033657\,i$ & $4.079915-0.033657\,i$ & $0$ \\
$6.0$ & $4.442678-0.033734\,i$ & $4.442678-0.033734\,i$ & $0$ & $4.444952-0.033713\,i$ & $4.444952-0.033713\,i$ & $0$ & $4.449512-0.033671\,i$ & $4.449512-0.033671\,i$ & $0$ \\
$6.5$ & $4.812903-0.033735\,i$ & $4.812903-0.033735\,i$ & $0$ & $4.815001-0.033717\,i$ & $4.815001-0.033717\,i$ & $0$ & $4.819208-0.033681\,i$ & $4.819208-0.033681\,i$ & $0$ \\
$7.0$ & $5.183127-0.033736\,i$ & $5.183127-0.033736\,i$ & $0$ & $5.185075-0.033720\,i$ & $5.185075-0.033720\,i$ & $0$ & $5.188979-0.033689\,i$ & $5.188979-0.033689\,i$ & $0$ \\
$7.5$ & $5.553351-0.033736\,i$ & $5.553351-0.033736\,i$ & $0$ & $5.555169-0.033723\,i$ & $5.555169-0.033723\,i$ & $0$ & $5.558811-0.033696\,i$ & $5.558811-0.033696\,i$ & $0$ \\
$8.0$ & $5.923575-0.033737\,i$ & $5.923575-0.033737\,i$ & $0$ & $5.925279-0.033725\,i$ & $5.925279-0.033725\,i$ & $0$ & $5.928693-0.033701\,i$ & $5.928693-0.033701\,i$ & $0$ \\
$8.5$ & $6.293799-0.033737\,i$ & $6.293799-0.033737\,i$ & $0$ & $6.295403-0.033727\,i$ & $6.295403-0.033727\,i$ & $0$ & $6.298615-0.033706\,i$ & $6.298615-0.033706\,i$ & $0$ \\
$9.0$ & $6.664023-0.033738\,i$ & $6.664023-0.033738\,i$ & $0$ & $6.665538-0.033728\,i$ & $6.665538-0.033728\,i$ & $0$ & $6.668570-0.033710\,i$ & $6.668570-0.033710\,i$ & $0$ \\
$9.5$ & $7.034247-0.033738\,i$ & $7.034247-0.033738\,i$ & $0$ & $7.035682-0.033730\,i$ & $7.035682-0.033730\,i$ & $0$ & $7.038554-0.033713\,i$ & $7.038554-0.033713\,i$ & $0$ \\
$10.0$ & $7.404471-0.033738\,i$ & $7.404471-0.033738\,i$ & $0$ & $7.405834-0.033731\,i$ & $7.405834-0.033731\,i$ & $0$ & $7.408562-0.033716\,i$ & $7.408562-0.033716\,i$ & $0$ \\
\hline\hline
\end{tabular}%
}
\end{table*}

\begin{table*}[t]
\caption{Fundamental ($n=0$) massive-scalar quasinormal frequencies for the moderately charged generalized Proca black-hole benchmark with $(\alpha,\beta,\lambda,c_1,M,Q)=(1,1,0.2,2,1.00,0.60)$, $\Lambda_{\rm eff}\simeq 8.31\times 10^{-3}$, $r_h\simeq 2.045$, and $r_c\simeq 17.781$. The three angular sectors $\ell=0,1,2$ are displayed side by side for the same set of field masses $\mu$. In each block, $\omega^{(16)}$ denotes the 16th-order WKB value with Pad\'e approximant $\tilde m=8$, $\omega^{(14)}$ denotes the 14th-order value with $\tilde m=7$, and $\Delta$ is the relative difference between the two approximants, in percent.}
\label{tab:moderately-charged-bh-wkb}
\centering
\scriptsize
\setlength{\tabcolsep}{3.5pt}
\renewcommand{\arraystretch}{1.05}
\resizebox{\textwidth}{!}{%
\begin{tabular}{c c c c c c c c c c}
\hline\hline
 & \multicolumn{3}{c}{$\ell=0$} & \multicolumn{3}{c}{$\ell=1$} & \multicolumn{3}{c}{$\ell=2$} \\
\cline{2-4}\cline{5-7}\cline{8-10}
$\mu$ & $\omega^{(16)}$ & $\omega^{(14)}$ & $\Delta(\%)$ & $\omega^{(16)}$ & $\omega^{(14)}$ & $\Delta(\%)$ & $\omega^{(16)}$ & $\omega^{(14)}$ & $\Delta(\%)$ \\
\hline
$0$ & $0.098672-0.088848\,i$ & $0.098702-0.088765\,i$ & $0.0666$ & $0.258882-0.080696\,i$ & $0.258882-0.080696\,i$ & $0.0002$ & $0.428777-0.079364\,i$ & $0.428777-0.079364\,i$ & $0$ \\
$0.5$ & $0.370734-0.032277\,i$ & $0.370633-0.032177\,i$ & $0.0382$ & $0.410572-0.033211\,i$ & $0.410571-0.033209\,i$ & $0.00035$ & $0.511542-0.055279\,i$ & $0.511543-0.055281\,i$ & $0.00027$ \\
$1.0$ & $0.741543-0.033379\,i$ & $0.741542-0.033379\,i$ & $0.00018$ & $0.756253-0.032392\,i$ & $0.756251-0.032391\,i$ & $0.00035$ & $0.791016-0.030865\,i$ & $0.791018-0.030859\,i$ & $0.00083$ \\
$1.5$ & $1.112368-0.033523\,i$ & $1.112368-0.033523\,i$ & $0$ & $1.121797-0.033132\,i$ & $1.121797-0.033132\,i$ & $0$ & $1.141760-0.032289\,i$ & $1.141760-0.032289\,i$ & $0$ \\
$2.0$ & $1.483185-0.033571\,i$ & $1.483185-0.033571\,i$ & $0$ & $1.490170-0.033358\,i$ & $1.490170-0.033358\,i$ & $0$ & $1.504561-0.032913\,i$ & $1.504561-0.032913\,i$ & $0$ \\
$2.5$ & $1.853999-0.033592\,i$ & $1.853999-0.033592\,i$ & $0$ & $1.859555-0.033458\,i$ & $1.859555-0.033458\,i$ & $0$ & $1.870877-0.033182\,i$ & $1.870877-0.033182\,i$ & $0$ \\
$3.0$ & $2.224810-0.033604\,i$ & $2.224810-0.033604\,i$ & $0$ & $2.229427-0.033511\,i$ & $2.229427-0.033511\,i$ & $0$ & $2.238778-0.033322\,i$ & $2.238778-0.033322\,i$ & $0$ \\
$3.5$ & $2.595620-0.033611\,i$ & $2.595620-0.033611\,i$ & $0$ & $2.599570-0.033543\,i$ & $2.599570-0.033543\,i$ & $0$ & $2.607544-0.033406\,i$ & $2.607544-0.033406\,i$ & $0$ \\
$4.0$ & $2.966429-0.033615\,i$ & $2.966429-0.033615\,i$ & $0$ & $2.969882-0.033564\,i$ & $2.969882-0.033564\,i$ & $0$ & $2.976835-0.033459\,i$ & $2.976835-0.033459\,i$ & $0$ \\
$4.5$ & $3.337237-0.033619\,i$ & $3.337237-0.033619\,i$ & $0$ & $3.340304-0.033578\,i$ & $3.340304-0.033578\,i$ & $0$ & $3.346471-0.033495\,i$ & $3.346471-0.033495\,i$ & $0$ \\
$5.0$ & $3.708045-0.033621\,i$ & $3.708045-0.033621\,i$ & $0$ & $3.710804-0.033588\,i$ & $3.710804-0.033588\,i$ & $0$ & $3.716345-0.033521\,i$ & $3.716345-0.033521\,i$ & $0$ \\
$5.5$ & $4.078853-0.033622\,i$ & $4.078853-0.033622\,i$ & $0$ & $4.081360-0.033595\,i$ & $4.081360-0.033595\,i$ & $0$ & $4.086391-0.033540\,i$ & $4.086391-0.033540\,i$ & $0$ \\
$6.0$ & $4.449660-0.033624\,i$ & $4.449660-0.033624\,i$ & $0$ & $4.451957-0.033601\,i$ & $4.451957-0.033601\,i$ & $0$ & $4.456566-0.033555\,i$ & $4.456566-0.033555\,i$ & $0$ \\
$6.5$ & $4.820468-0.033625\,i$ & $4.820468-0.033625\,i$ & $0$ & $4.822587-0.033605\,i$ & $4.822587-0.033605\,i$ & $0$ & $4.826838-0.033566\,i$ & $4.826838-0.033566\,i$ & $0$ \\
$7.0$ & $5.191275-0.033625\,i$ & $5.191275-0.033625\,i$ & $0$ & $5.193243-0.033609\,i$ & $5.193243-0.033609\,i$ & $0$ & $5.197188-0.033575\,i$ & $5.197188-0.033575\,i$ & $0$ \\
$7.5$ & $5.562082-0.033626\,i$ & $5.562082-0.033626\,i$ & $0$ & $5.563918-0.033611\,i$ & $5.563918-0.033611\,i$ & $0$ & $5.567599-0.033582\,i$ & $5.567599-0.033582\,i$ & $0$ \\
$8.0$ & $5.932889-0.033626\,i$ & $5.932889-0.033626\,i$ & $0$ & $5.934610-0.033614\,i$ & $5.934610-0.033614\,i$ & $0$ & $5.938059-0.033588\,i$ & $5.938059-0.033588\,i$ & $0$ \\
$8.5$ & $6.303695-0.033627\,i$ & $6.303695-0.033627\,i$ & $0$ & $6.305315-0.033616\,i$ & $6.305315-0.033616\,i$ & $0$ & $6.308561-0.033593\,i$ & $6.308561-0.033593\,i$ & $0$ \\
$9.0$ & $6.674502-0.033627\,i$ & $6.674502-0.033627\,i$ & $0$ & $6.676032-0.033617\,i$ & $6.676032-0.033617\,i$ & $0$ & $6.679096-0.033597\,i$ & $6.679096-0.033597\,i$ & $0$ \\
$9.5$ & $7.045309-0.033628\,i$ & $7.045309-0.033628\,i$ & $0$ & $7.046758-0.033618\,i$ & $7.046758-0.033618\,i$ & $0$ & $7.049660-0.033600\,i$ & $7.049660-0.033600\,i$ & $0$ \\
$10.0$ & $7.416115-0.033628\,i$ & $7.416115-0.033628\,i$ & $0$ & $7.417492-0.033620\,i$ & $7.417492-0.033620\,i$ & $0$ & $7.420249-0.033603\,i$ & $7.420249-0.033603\,i$ & $0$ \\
\hline\hline
\end{tabular}%
}
\end{table*}

\begin{table*}[t]
\caption{Fundamental ($n=0$) massive-scalar quasinormal frequencies for the strongly charged generalized Proca black-hole benchmark with $(\alpha,\beta,\lambda,c_1,M,Q)=(1,1,0.2,2,1.00,0.90)$, $\Lambda_{\rm eff}\simeq 8.31\times 10^{-3}$, $r_h\simeq 1.736$, and $r_c\simeq 17.783$. The three angular sectors $\ell=0,1,2$ are displayed side by side for the same set of field masses $\mu$. In each block, $\omega^{(16)}$ denotes the 16th-order WKB value with Pad\'e approximant $\tilde m=8$, $\omega^{(14)}$ denotes the 14th-order value with $\tilde m=7$, and $\Delta$ is the relative difference between the two approximants, in percent.}
\label{tab:strongly-charged-bh-wkb}
\centering
\scriptsize
\setlength{\tabcolsep}{3.5pt}
\renewcommand{\arraystretch}{1.05}
\resizebox{\textwidth}{!}{%
\begin{tabular}{c c c c c c c c c c}
\hline\hline
 & \multicolumn{3}{c}{$\ell=0$} & \multicolumn{3}{c}{$\ell=1$} & \multicolumn{3}{c}{$\ell=2$} \\
\cline{2-4}\cline{5-7}\cline{8-10}
$\mu$ & $\omega^{(16)}$ & $\omega^{(14)}$ & $\Delta(\%)$ & $\omega^{(16)}$ & $\omega^{(14)}$ & $\Delta(\%)$ & $\omega^{(16)}$ & $\omega^{(14)}$ & $\Delta(\%)$ \\
\hline
$0$ & $0.100942-0.079304\,i$ & $0.100942-0.079304\,i$ & $0.0006$ & $0.270639-0.072807\,i$ & $0.270639-0.072806\,i$ & $0.0003$ & $0.448143-0.071755\,i$ & $0.448143-0.071755\,i$ & $0$ \\
$0.5$ & $0.371053-0.032114\,i$ & $0.370983-0.032065\,i$ & $0.0230$ & $0.411876-0.031912\,i$ & $0.411878-0.031925\,i$ & $0.00331$ & $0.522267-0.053083\,i$ & $0.522267-0.053082\,i$ & $0.00021$ \\
$1.0$ & $0.742356-0.033250\,i$ & $0.742352-0.033252\,i$ & $0.00053$ & $0.757186-0.032176\,i$ & $0.757186-0.032177\,i$ & $0.0001$ & $0.792504-0.030128\,i$ & $0.792510-0.030120\,i$ & $0.00128$ \\
$1.5$ & $1.113622-0.033389\,i$ & $1.113622-0.033389\,i$ & $0$ & $1.123135-0.032964\,i$ & $1.123135-0.032964\,i$ & $0$ & $1.143316-0.032022\,i$ & $1.143316-0.032022\,i$ & $0$ \\
$2.0$ & $1.484876-0.033435\,i$ & $1.484876-0.033435\,i$ & $0$ & $1.491924-0.033204\,i$ & $1.491924-0.033204\,i$ & $0$ & $1.506462-0.032714\,i$ & $1.506462-0.032714\,i$ & $0$ \\
$2.5$ & $1.856123-0.033456\,i$ & $1.856123-0.033456\,i$ & $0$ & $1.861730-0.033310\,i$ & $1.861730-0.033310\,i$ & $0$ & $1.873163-0.033007\,i$ & $1.873163-0.033007\,i$ & $0$ \\
$3.0$ & $2.227366-0.033467\,i$ & $2.227366-0.033467\,i$ & $0$ & $2.232025-0.033366\,i$ & $2.232025-0.033366\,i$ & $0$ & $2.241467-0.033160\,i$ & $2.241467-0.033160\,i$ & $0$ \\
$3.5$ & $2.598606-0.033474\,i$ & $2.598606-0.033474\,i$ & $0$ & $2.602593-0.033400\,i$ & $2.602593-0.033400\,i$ & $0$ & $2.610643-0.033250\,i$ & $2.610643-0.033250\,i$ & $0$ \\
$4.0$ & $2.969846-0.033478\,i$ & $2.969846-0.033478\,i$ & $0$ & $2.973330-0.033422\,i$ & $2.973330-0.033422\,i$ & $0$ & $2.980350-0.033308\,i$ & $2.980350-0.033308\,i$ & $0$ \\
$4.5$ & $3.341084-0.033481\,i$ & $3.341084-0.033481\,i$ & $0$ & $3.344179-0.033437\,i$ & $3.344179-0.033437\,i$ & $0$ & $3.350404-0.033347\,i$ & $3.350404-0.033347\,i$ & $0$ \\
$5.0$ & $3.712321-0.033483\,i$ & $3.712321-0.033483\,i$ & $0$ & $3.715105-0.033447\,i$ & $3.715105-0.033447\,i$ & $0$ & $3.720699-0.033375\,i$ & $3.720699-0.033375\,i$ & $0$ \\
$5.5$ & $4.083559-0.033485\,i$ & $4.083559-0.033485\,i$ & $0$ & $4.086088-0.033455\,i$ & $4.086088-0.033455\,i$ & $0$ & $4.091167-0.033396\,i$ & $4.091167-0.033396\,i$ & $0$ \\
$6.0$ & $4.454795-0.033486\,i$ & $4.454795-0.033486\,i$ & $0$ & $4.457114-0.033461\,i$ & $4.457114-0.033461\,i$ & $0$ & $4.461765-0.033411\,i$ & $4.461765-0.033411\,i$ & $0$ \\
$6.5$ & $4.826032-0.033487\,i$ & $4.826032-0.033487\,i$ & $0$ & $4.828171-0.033466\,i$ & $4.828171-0.033466\,i$ & $0$ & $4.832462-0.033423\,i$ & $4.832462-0.033423\,i$ & $0$ \\
$7.0$ & $5.197268-0.033488\,i$ & $5.197268-0.033488\,i$ & $0$ & $5.199254-0.033469\,i$ & $5.199254-0.033469\,i$ & $0$ & $5.203236-0.033433\,i$ & $5.203236-0.033433\,i$ & $0$ \\
$7.5$ & $5.568504-0.033488\,i$ & $5.568504-0.033488\,i$ & $0$ & $5.570357-0.033472\,i$ & $5.570357-0.033472\,i$ & $0$ & $5.574072-0.033441\,i$ & $5.574072-0.033441\,i$ & $0$ \\
$8.0$ & $5.939740-0.033489\,i$ & $5.939740-0.033489\,i$ & $0$ & $5.941477-0.033475\,i$ & $5.941477-0.033475\,i$ & $0$ & $5.944958-0.033447\,i$ & $5.944958-0.033447\,i$ & $0$ \\
$8.5$ & $6.310975-0.033489\,i$ & $6.310975-0.033489\,i$ & $0$ & $6.312610-0.033477\,i$ & $6.312610-0.033477\,i$ & $0$ & $6.315886-0.033452\,i$ & $6.315886-0.033452\,i$ & $0$ \\
$9.0$ & $6.682211-0.033489\,i$ & $6.682211-0.033489\,i$ & $0$ & $6.683755-0.033478\,i$ & $6.683755-0.033478\,i$ & $0$ & $6.686848-0.033456\,i$ & $6.686848-0.033456\,i$ & $0$ \\
$9.5$ & $7.053446-0.033490\,i$ & $7.053446-0.033490\,i$ & $0$ & $7.054909-0.033480\,i$ & $7.054909-0.033480\,i$ & $0$ & $7.057838-0.033460\,i$ & $7.057838-0.033460\,i$ & $0$ \\
$10.0$ & $7.424682-0.033490\,i$ & $7.424682-0.033490\,i$ & $0$ & $7.426071-0.033481\,i$ & $7.426071-0.033481\,i$ & $0$ & $7.428854-0.033463\,i$ & $7.428854-0.033463\,i$ & $0$ \\
\hline\hline
\end{tabular}%
}
\end{table*}

\begin{table*}[t]
\caption{Fundamental ($n=0$) massive-scalar quasinormal frequencies for the large neutral generalized Proca black-hole benchmark with $(\alpha,\beta,\lambda,c_1,M,Q)=(1,1,0.2,2,3.20,0)$, $\Lambda_{\rm eff}\simeq 8.31\times 10^{-3}$, $r_h\simeq 9.423$, and $r_c\simeq 12.448$. The three angular sectors $\ell=0,1,2$ are displayed side by side for the same set of field masses $\mu$. In each block, $\omega^{(16)}$ denotes the 16th-order WKB value with Pad\'e approximant $\tilde m=8$, $\omega^{(14)}$ denotes the 14th-order value with $\tilde m=7$, and $\Delta$ is the relative difference between the two approximants, in percent.}
\label{tab:large-neutral-bh-wkb}
\centering
\scriptsize
\setlength{\tabcolsep}{3.5pt}
\renewcommand{\arraystretch}{1.05}
\resizebox{\textwidth}{!}{%
\begin{tabular}{c c c c c c c c c c}
\hline\hline
 & \multicolumn{3}{c}{$\ell=0$} & \multicolumn{3}{c}{$\ell=1$} & \multicolumn{3}{c}{$\ell=2$} \\
\cline{2-4}\cline{5-7}\cline{8-10}
$\mu$ & $\omega^{(16)}$ & $\omega^{(14)}$ & $\Delta(\%)$ & $\omega^{(16)}$ & $\omega^{(14)}$ & $\Delta(\%)$ & $\omega^{(16)}$ & $\omega^{(14)}$ & $\Delta(\%)$ \\
\hline
$0$ & $0.001950-0.010976\,i$ & $0.001962-0.010908\,i$ & $0.618$ & $0.017085-0.006507\,i$ & $0.017085-0.006508\,i$ & $0.002$ & $0.030825-0.006441\,i$ & $0.030825-0.006441\,i$ & $0$ \\
$0.5$ & $0.068705-0.006285\,i$ & $0.068705-0.006285\,i$ & $0$ & $0.071019-0.006283\,i$ & $0.071019-0.006283\,i$ & $0$ & $0.075447-0.006282\,i$ & $0.075447-0.006282\,i$ & $0$ \\
$1.0$ & $0.137833-0.006287\,i$ & $0.137833-0.006287\,i$ & $0$ & $0.138999-0.006286\,i$ & $0.138999-0.006286\,i$ & $0$ & $0.141304-0.006284\,i$ & $0.141304-0.006284\,i$ & $0$ \\
$1.5$ & $0.206867-0.006287\,i$ & $0.206867-0.006287\,i$ & $0$ & $0.207645-0.006287\,i$ & $0.207645-0.006287\,i$ & $0$ & $0.209194-0.006286\,i$ & $0.209194-0.006286\,i$ & $0$ \\
$2.0$ & $0.275878-0.006287\,i$ & $0.275878-0.006287\,i$ & $0$ & $0.276462-0.006287\,i$ & $0.276462-0.006287\,i$ & $0$ & $0.277626-0.006286\,i$ & $0.277626-0.006286\,i$ & $0$ \\
$2.5$ & $0.344879-0.006287\,i$ & $0.344879-0.006287\,i$ & $0$ & $0.345346-0.006287\,i$ & $0.345346-0.006287\,i$ & $0$ & $0.346279-0.006287\,i$ & $0.346279-0.006287\,i$ & $0$ \\
$3.0$ & $0.413876-0.006287\,i$ & $0.413876-0.006287\,i$ & $0$ & $0.414265-0.006287\,i$ & $0.414265-0.006287\,i$ & $0$ & $0.415043-0.006287\,i$ & $0.415043-0.006287\,i$ & $0$ \\
$3.5$ & $0.482869-0.006287\,i$ & $0.482869-0.006287\,i$ & $0$ & $0.483203-0.006287\,i$ & $0.483203-0.006287\,i$ & $0$ & $0.483870-0.006287\,i$ & $0.483870-0.006287\,i$ & $0$ \\
$4.0$ & $0.551861-0.006287\,i$ & $0.551861-0.006287\,i$ & $0$ & $0.552154-0.006287\,i$ & $0.552154-0.006287\,i$ & $0$ & $0.552737-0.006287\,i$ & $0.552737-0.006287\,i$ & $0$ \\
$4.5$ & $0.620852-0.006287\,i$ & $0.620852-0.006287\,i$ & $0$ & $0.621112-0.006287\,i$ & $0.621112-0.006287\,i$ & $0$ & $0.621631-0.006287\,i$ & $0.621631-0.006287\,i$ & $0$ \\
$5.0$ & $0.689843-0.006287\,i$ & $0.689843-0.006287\,i$ & $0$ & $0.690076-0.006287\,i$ & $0.690076-0.006287\,i$ & $0$ & $0.690544-0.006287\,i$ & $0.690544-0.006287\,i$ & $0$ \\
$5.5$ & $0.758832-0.006287\,i$ & $0.758832-0.006287\,i$ & $0$ & $0.759045-0.006287\,i$ & $0.759045-0.006287\,i$ & $0$ & $0.759470-0.006287\,i$ & $0.759470-0.006287\,i$ & $0$ \\
$6.0$ & $0.827822-0.006287\,i$ & $0.827822-0.006287\,i$ & $0$ & $0.828016-0.006287\,i$ & $0.828016-0.006287\,i$ & $0$ & $0.828406-0.006287\,i$ & $0.828406-0.006287\,i$ & $0$ \\
$6.5$ & $0.896810-0.006287\,i$ & $0.896810-0.006287\,i$ & $0$ & $0.896990-0.006287\,i$ & $0.896990-0.006287\,i$ & $0$ & $0.897350-0.006287\,i$ & $0.897350-0.006287\,i$ & $0$ \\
$7.0$ & $0.965799-0.006287\,i$ & $0.965799-0.006287\,i$ & $0$ & $0.965966-0.006287\,i$ & $0.965966-0.006287\,i$ & $0$ & $0.966300-0.006287\,i$ & $0.966300-0.006287\,i$ & $0$ \\
$7.5$ & $1.034788-0.006287\,i$ & $1.034788-0.006287\,i$ & $0$ & $1.034943-0.006287\,i$ & $1.034943-0.006287\,i$ & $0$ & $1.035255-0.006287\,i$ & $1.035255-0.006287\,i$ & $0$ \\
$8.0$ & $1.103776-0.006287\,i$ & $1.103776-0.006287\,i$ & $0$ & $1.103922-0.006287\,i$ & $1.103922-0.006287\,i$ & $0$ & $1.104214-0.006287\,i$ & $1.104214-0.006287\,i$ & $0$ \\
$8.5$ & $1.172764-0.006287\,i$ & $1.172764-0.006287\,i$ & $0$ & $1.172901-0.006287\,i$ & $1.172901-0.006287\,i$ & $0$ & $1.173176-0.006287\,i$ & $1.173176-0.006287\,i$ & $0$ \\
$9.0$ & $1.241752-0.006287\,i$ & $1.241752-0.006287\,i$ & $0$ & $1.241882-0.006287\,i$ & $1.241882-0.006287\,i$ & $0$ & $1.242141-0.006287\,i$ & $1.242141-0.006287\,i$ & $0$ \\
$9.5$ & $1.310740-0.006287\,i$ & $1.310740-0.006287\,i$ & $0$ & $1.310863-0.006287\,i$ & $1.310863-0.006287\,i$ & $0$ & $1.311109-0.006287\,i$ & $1.311109-0.006287\,i$ & $0$ \\
$10.0$ & $1.379728-0.006287\,i$ & $1.379728-0.006287\,i$ & $0$ & $1.379845-0.006287\,i$ & $1.379845-0.006287\,i$ & $0$ & $1.380078-0.006287\,i$ & $1.380078-0.006287\,i$ & $0$ \\
\hline\hline
\end{tabular}%
}
\end{table*}

\begin{table*}[t]
\caption{First three overtones ($n=1,2,3$) of the massive-scalar quasinormal frequencies for the reference neutral generalized Proca black-hole with $(\alpha,\beta,\lambda,c_1,M,Q,\ell)=(1,1,0.2,2,1.00,0,1)$, $\Lambda_{\rm eff}\simeq 8.31\times 10^{-3}$, $r_h\simeq 2.248$, and $r_c\simeq 17.776$. The $\ell=1$ sector is displayed for the same set of field masses $\mu$, with the overtones $n=1,2,3$ shown side by side. In each block, $\omega^{(16)}$ denotes the 16th-order WKB value with Pad\'e approximant $\tilde m=8$, $\omega^{(14)}$ denotes the 14th-order value with $\tilde m=7$, and $\Delta$ is the relative difference between the two approximants, in percent.}
\label{tab:reference-neutral-bh-wkb-overtones}
\centering
\scriptsize
\setlength{\tabcolsep}{3.5pt}
\renewcommand{\arraystretch}{1.05}
\resizebox{\textwidth}{!}{%
\begin{tabular}{c c c c c c c c c c}
\hline\hline
 & \multicolumn{3}{c}{$n=1$} & \multicolumn{3}{c}{$n=2$} & \multicolumn{3}{c}{$n=3$} \\
\cline{2-4}\cline{5-7}\cline{8-10}
$\mu$ & $\omega^{(16)}$ & $\omega^{(14)}$ & $\Delta(\%)$ & $\omega^{(16)}$ & $\omega^{(14)}$ & $\Delta(\%)$ & $\omega^{(16)}$ & $\omega^{(14)}$ & $\Delta(\%)$ \\
\hline
$0$ & $0.228643-0.262594\,i$ & $0.228643-0.262595\,i$ & $0.0002$ & $0.199840-0.459072\,i$ & $0.199839-0.459073\,i$ & $0.00035$ & $0.177412-0.668861\,i$ & $0.176535-0.669600\,i$ & $0.166$ \\
$0.5$ & $0.412070-0.100542\,i$ & $0.411898-0.100467\,i$ & $0.0442$ & $0.417511-0.170555\,i$ & $0.418017-0.172137\,i$ & $0.368$ & $0.423256-0.241657\,i$ & $0.423298-0.241688\,i$ & $0.0106$ \\
$1.0$ & $0.756541-0.097735\,i$ & $0.756537-0.097736\,i$ & $0.00051$ & $0.759419-0.162863\,i$ & $0.759378-0.163014\,i$ & $0.0201$ & $0.761498-0.227627\,i$ & $0.763370-0.229073\,i$ & $0.298$ \\
$1.5$ & $1.120877-0.099858\,i$ & $1.120877-0.099858\,i$ & $0$ & $1.122650-0.166574\,i$ & $1.122650-0.166574\,i$ & $0.00002$ & $1.125265-0.233495\,i$ & $1.125268-0.233502\,i$ & $0.00064$ \\
$2.0$ & $1.488434-0.100479\,i$ & $1.488434-0.100479\,i$ & $0$ & $1.489716-0.167553\,i$ & $1.489716-0.167553\,i$ & $0$ & $1.491620-0.234757\,i$ & $1.491620-0.234757\,i$ & $0$ \\
$2.5$ & $1.857111-0.100753\,i$ & $1.857111-0.100753\,i$ & $0$ & $1.858123-0.167979\,i$ & $1.858123-0.167979\,i$ & $0$ & $1.859631-0.235290\,i$ & $1.859631-0.235290\,i$ & $0$ \\
$3.0$ & $2.226319-0.100898\,i$ & $2.226319-0.100898\,i$ & $0$ & $2.227157-0.168204\,i$ & $2.227157-0.168204\,i$ & $0$ & $2.228408-0.235570\,i$ & $2.228408-0.235570\,i$ & $0$ \\
$3.5$ & $2.595823-0.100985\,i$ & $2.595823-0.100985\,i$ & $0$ & $2.596539-0.168338\,i$ & $2.596539-0.168338\,i$ & $0$ & $2.597609-0.235736\,i$ & $2.597609-0.235736\,i$ & $0$ \\
$4.0$ & $2.965510-0.101041\,i$ & $2.965510-0.101041\,i$ & $0$ & $2.966135-0.168425\,i$ & $2.966135-0.168425\,i$ & $0$ & $2.967070-0.235842\,i$ & $2.967070-0.235842\,i$ & $0$ \\
$4.5$ & $3.335317-0.101079\,i$ & $3.335317-0.101079\,i$ & $0$ & $3.335872-0.168484\,i$ & $3.335872-0.168484\,i$ & $0$ & $3.336702-0.235915\,i$ & $3.336702-0.235915\,i$ & $0$ \\
$5.0$ & $3.705209-0.101107\,i$ & $3.705209-0.101107\,i$ & $0$ & $3.705708-0.168526\,i$ & $3.705708-0.168526\,i$ & $0$ & $3.706454-0.235967\,i$ & $3.706454-0.235967\,i$ & $0$ \\
$5.5$ & $4.075161-0.101127\,i$ & $4.075161-0.101127\,i$ & $0$ & $4.075614-0.168557\,i$ & $4.075614-0.168557\,i$ & $0$ & $4.076293-0.236005\,i$ & $4.076293-0.236005\,i$ & $0$ \\
$6.0$ & $4.445159-0.101142\,i$ & $4.445159-0.101142\,i$ & $0$ & $4.445574-0.168580\,i$ & $4.445574-0.168580\,i$ & $0$ & $4.446196-0.236034\,i$ & $4.446196-0.236034\,i$ & $0$ \\
$6.5$ & $4.815192-0.101154\,i$ & $4.815192-0.101154\,i$ & $0$ & $4.815575-0.168599\,i$ & $4.815575-0.168599\,i$ & $0$ & $4.816149-0.236056\,i$ & $4.816149-0.236056\,i$ & $0$ \\
$7.0$ & $5.185253-0.101163\,i$ & $5.185253-0.101163\,i$ & $0$ & $5.185608-0.168613\,i$ & $5.185608-0.168613\,i$ & $0$ & $5.186141-0.236074\,i$ & $5.186141-0.236074\,i$ & $0$ \\
$7.5$ & $5.555335-0.101171\,i$ & $5.555335-0.101171\,i$ & $0$ & $5.555666-0.168625\,i$ & $5.555666-0.168625\,i$ & $0$ & $5.556163-0.236088\,i$ & $5.556163-0.236088\,i$ & $0$ \\
$8.0$ & $5.925435-0.101177\,i$ & $5.925435-0.101177\,i$ & $0$ & $5.925745-0.168634\,i$ & $5.925745-0.168634\,i$ & $0$ & $5.926211-0.236100\,i$ & $5.926211-0.236100\,i$ & $0$ \\
$8.5$ & $6.295549-0.101182\,i$ & $6.295549-0.101182\,i$ & $0$ & $6.295842-0.168642\,i$ & $6.295842-0.168642\,i$ & $0$ & $6.296280-0.236110\,i$ & $6.296280-0.236110\,i$ & $0$ \\
$9.0$ & $6.665676-0.101187\,i$ & $6.665676-0.101187\,i$ & $0$ & $6.665952-0.168649\,i$ & $6.665952-0.168649\,i$ & $0$ & $6.666366-0.236118\,i$ & $6.666366-0.236118\,i$ & $0$ \\
$9.5$ & $7.035813-0.101190\,i$ & $7.035813-0.101190\,i$ & $0$ & $7.036074-0.168655\,i$ & $7.036074-0.168655\,i$ & $0$ & $7.036467-0.236125\,i$ & $7.036467-0.236125\,i$ & $0$ \\
$10.0$ & $7.405958-0.101193\,i$ & $7.405958-0.101193\,i$ & $0$ & $7.406207-0.168659\,i$ & $7.406207-0.168659\,i$ & $0$ & $7.406580-0.236131\,i$ & $7.406580-0.236131\,i$ & $0$ \\
\hline\hline
\end{tabular}%
}
\end{table*}

Tables~\ref{tab:small-neutral-bh-wkb}--\ref{tab:large-neutral-bh-wkb} show three robust effects. First, at fixed background the field mass primarily raises the oscillation frequency and rapidly lowers the damping until a background-dependent plateau is reached. In the reference neutral case, for example, the $\ell=1$ fundamental evolves from $\omega\simeq 0.249251-0.084986\,i$ at $\mu=0$ to $0.409030-0.034093\,i$ at $\mu=0.5$, and by $\mu=10$ it has reached $7.405834-0.033731\,i$. Second, enlarging the black hole strongly decreases the ringing frequency: the massless $\ell=1$ mode changes from $0.521864-0.173755\,i$ for the small neutral case to $0.249251-0.084986\,i$ for the reference case and to only $0.017085-0.006507\,i$ for the large near-Nariai-like benchmark. Third, at fixed $M=1$ the electric charge slightly increases ${\rm Re}\,\omega$ and decreases $|{\rm Im}\,\omega|$; for $\ell=1$ and $\mu=0$ one finds $0.249251-0.084986\,i$ at $Q=0$, $0.258882-0.080696\,i$ at $Q=0.6$, and $0.270639-0.072807\,i$ at $Q=0.9$, with the same tendency still visible at $\mu=0.5$.

The angular-momentum splitting is appreciable only for light fields. In the reference neutral black-hole case, the low-mass fundamental mode moves from $0.094568-0.094048\,i$ for $\ell=0$ to $0.249251-0.084986\,i$ for $\ell=1$ and $0.413699-0.083425\,i$ for $\ell=2$, whereas at $\mu=10$ the same three entries compress to $7.404471-0.033738\,i$, $7.405834-0.033731\,i$, and $7.408562-0.033716\,i$. The large neutral benchmark is even more striking: at $\mu=10$ all three sectors share essentially the same damping, ${\rm Im}\,\omega\simeq -6.287\times 10^{-3}$. This near-degeneracy already points to the large-$\mu$ asymptotics derived below, in which the leading damping becomes independent of both $\mu$ and $\ell$. The Pad\'e comparison also indicates that the WKB data are numerically stable: aside from a few low-mass entries, $\Delta$ is typically tiny and never exceeds $0.618\%$ in the fundamental tables.

The reference neutral background also provides a clean overtone behavior, summarized in Tables~\ref{tab:reference-neutral-bh-wkb} and \ref{tab:reference-neutral-bh-wkb-overtones}. At fixed $\ell=1$, increasing $n$ barely shifts the real part once $\mu$ is moderate or large, while the damping grows almost exactly according to the expected $(n+\frac{1}{2})$ pattern. For instance, at $\mu=10$ the $n=0,1,2,3$ frequencies are $7.405834-0.033731\,i$, $7.405958-0.101193\,i$, $7.406207-0.168659\,i$, and $7.406580-0.236131\,i$, so the imaginary parts are already extremely close to the ratio $1:3:5:7$. This makes the reference neutral case a convenient bridge between the tabulated WKB spectrum and the time-domain Prony extraction discussed next.

\clearpage

\begin{figure}[t]
\centering
\includegraphics[width=0.98\linewidth]{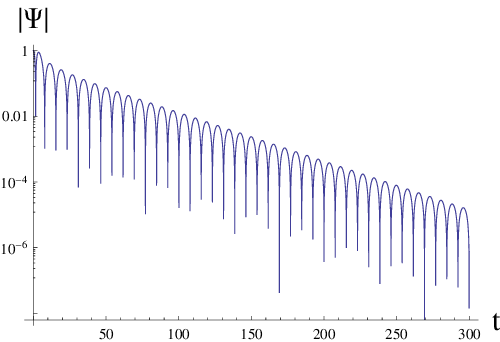}
\caption{Time-domain profile for the $\ell=1$ massive-scalar perturbation of the reference neutral generalized Proca black-hole benchmark with $(\alpha,\beta,\lambda,c_1,M,Q,\mu)=(1,1,0.2,2,1.00,0,0.5)$. A three-mode Prony fit yields $\omega_0\simeq 0.409022-0.034096\,i$, $\omega_1\simeq 0.411894-0.100556\,i$, and $\omega_2\simeq 0.416926-0.170631\,i$. In the de Sitter case the quasinormal ringing governs the decay throughout the evolution, without contamination from asymptotic tails, so the first few overtones can be extracted accurately. Comparing with the tabulated WKB values for the corresponding $n=0,1,2$ modes, namely $0.409030-0.034093\,i$, $0.412070-0.100542\,i$, and $0.417511-0.170555\,i$, the relative differences remain below about $0.15\%$ in the real part and below about $0.05\%$ in the imaginary part.}
\label{fig:td-reference-neutral-mu05}
\end{figure}

For the same reference-neutral type-II background at $\mu=0.5$, the Prony extraction shown in Fig.~\ref{fig:td-reference-neutral-mu05} is in excellent agreement with the WKB values listed in Table~\ref{tab:reference-neutral-bh-wkb} and Table~\ref{tab:reference-neutral-bh-wkb-overtones}. In particular, the three frequencies $0.409022-0.034096\,i$, $0.411894-0.100556\,i$, and $0.416926-0.170631\,i$ match the tabulated $n=0$, $n=1$, and $n=2$ values $0.409030-0.034093\,i$, $0.412070-0.100542\,i$, and $0.417511-0.170555\,i$, respectively. The relative discrepancies stay below about $0.15\%$ in the real part and below about $0.05\%$ in the imaginary part, which supports the interpretation that the first three oscillatory contributions seen in the waveform belong to the same mode sequence captured by the WKB tables.

A simple pattern visible in Tables~\ref{tab:small-neutral-bh-wkb}--\ref{tab:large-neutral-bh-wkb} is that, for sufficiently large $\mu$, the damping rate approaches a constant which is essentially independent of $\ell$ for each fixed black-hole background, while the real part grows almost linearly with $\mu$. This can be understood directly from the effective potential~\eqref{eq:potential}. In the regime
\begin{equation}
\mu^2 \gg \frac{\ell(\ell+1)}{r^2},
\qquad
\mu^2 \gg \left|\frac{f'(r)}{r}\right|,
\end{equation}
one has
\begin{equation}
\begin{split}
V(r)=&\,\mu^2 f(r)+f(r)\left[\frac{\ell(\ell+1)}{r^2}+\frac{f'(r)}{r}\right]\\
=&\,\mu^2 f(r)+\mathcal{O}(1).
\end{split}
\end{equation}
Therefore the barrier peak is determined, to leading order, by the maximum of $f(r)$ itself. Let $r_0$ be the point in the static region where $f'(r_0)=0$ and $f''(r_0)<0$, and denote $f_0\equiv f(r_0)$. Then the peak position satisfies $r_{\max}=r_0+\mathcal{O}(\mu^{-2})$, while
\begin{equation}
V_0=\mu^2 f_0+\mathcal{O}(1),
\qquad
V_0''=\mu^2 f_0^2 f''(r_0)+\mathcal{O}(1),
\end{equation}
where $V_0''$ is the second derivative with respect to the tortoise coordinate. Substituting these expressions into the leading WKB formula yields
\begin{equation}
\omega_{n\ell}^2=\mu^2 f_0-i\left(n+\frac{1}{2}\right)\mu f_0\sqrt{-2f''(r_0)}+\mathcal{O}(1),
\label{eq:large_mu_wkb_squared}
\end{equation}
and hence the simple large-mass asymptotic relation
\begin{equation}
\omega_{n\ell}=\mu\sqrt{f_0}-i\left(n+\frac{1}{2}\right)\sqrt{-\frac{f_0 f''(r_0)}{2}}+\mathcal{O}(\mu^{-1}).
\label{eq:large_mu_wkb}
\end{equation}
For the generalized Proca metric it is convenient to introduce
\begin{equation}
\begin{split}
m&\equiv M-Q,
\qquad
C\equiv Q+\frac{\lambda(M-Q)}{2\beta},\\
\Delta_0&\equiv\sqrt{B+\frac{8\alpha C}{r_0^3}},
\end{split}
\end{equation}
where $r_0$ is the point at which $f(r)$ reaches its maximum in the static region. The condition $f'(r_0)=0$ can then be written compactly as
\begin{equation}
r_0^3\bigl(A-\Delta_0\bigr)+2\alpha\left(m+\frac{3C}{\Delta_0}\right)=0.
\label{eq:r0_condition}
\end{equation}
Using this relation, the two quantities entering Eq.~\eqref{eq:large_mu_wkb} simplify to
\begin{equation}
\begin{split}
f_0&=1-\frac{3}{r_0}\left(m+\frac{C}{\Delta_0}\right),\\
f''(r_0)&=-\frac{3}{r_0^3}\left(2m-\frac{C}{\Delta_0}+\frac{3BC}{\Delta_0^3}\right).
\end{split}
\end{equation}
Therefore the large-mass quasinormal frequency can be written in the compact form
\begin{equation}
\omega_{n\ell}=\mu\,\Omega_R-i\left(n+\frac{1}{2}\right)\Omega_I+\mathcal{O}(\mu^{-1}),
\label{eq:large_mu_compact}
\end{equation}
with
\begin{equation}
\Omega_R\equiv\sqrt{1-\frac{3}{r_0}\left(m+\frac{C}{\Delta_0}\right)},
\end{equation}
\begin{equation}
\Omega_I\equiv\sqrt{\frac{3}{2r_0^3}\left(1-\frac{3}{r_0}\left(m+\frac{C}{\Delta_0}\right)\right)
\left(2m-\frac{C}{\Delta_0}+\frac{3BC}{\Delta_0^3}\right)}.
\end{equation}
This is essentially the simplest useful expression: all dependence on $(M,Q,\alpha,\beta,c_1,\lambda)$ is encoded through $A$, $B$, $C$, and the single geometric point $r_0$ determined by Eq.~\eqref{eq:r0_condition}. Thus the leading damping rate is independent of both $\mu$ and $\ell$, whereas the angular-momentum dependence first appears only in the subleading $\mathcal{O}(\mu^{-1})$ corrections. This explains why, in Tables~I--V, the imaginary part settles to an almost constant value for large $\mu$, and why the frequencies for different $\ell$ become nearly degenerate in that regime. The fact that the constant differs from one benchmark to another simply reflects that each background has its own values of $r_0$ and $\Delta_0$, i.e. its own barrier height and curvature. The same asymptotic picture also governs Table~\ref{tab:reference-neutral-bh-wkb-overtones}: at fixed background and fixed $\ell=1$, the large-$\mu$ real part remains almost unchanged as $n$ varies, while the limiting damping rates follow the expected $(n+\frac{1}{2})$ scaling and therefore approach the approximate ratio $3:5:7$ across the $n=1,2,3$ overtones shown there. A direct check against the $\,\mu=10$, $\ell=2$, $n=0$ entries in Tables~\ref{tab:small-neutral-bh-wkb}--\ref{tab:large-neutral-bh-wkb} shows that Eq.~\eqref{eq:large_mu_compact} reproduces all five benchmarks with relative errors below $0.1\%$ in the real part and below $0.25\%$ in the imaginary part.

A final remark concerns the strong-cosmic-censorship diagnostic
$\beta_{\rm SCC}\equiv \alpha_{\rm gap}/\kappa_-$, where
$\alpha_{\rm gap}=\min(-{\rm Im}\,\omega)$ and $\kappa_-$ is the surface gravity of the inner Cauchy horizon. The five benchmark points listed in Tables~\ref{tab:small-neutral-bh-wkb}--\ref{tab:large-neutral-bh-wkb} have only an event horizon and a cosmological horizon, so $\beta_{\rm SCC}$ is not defined for them directly. Nevertheless, for the same couplings $(\alpha,\beta,\lambda,c_1)=(1,1,0.2,2)$ the generalized Proca metric does admit a three-horizon sector once the Proca hair parameter $Q$ is increased, and then the de Sitter branch described by Eqs.~\eqref{eq:ds_exact_qnm}--\eqref{eq:ds_heavy_qnm} becomes directly relevant to the SCC question.

In the small-black-hole regime the answer is especially clear. For the light-field de Sitter family, the $n=0$, $\ell=0$, minus-solution of Eq.~\eqref{eq:ds_exact_qnm} has damping
\begin{equation}
\begin{split}
\alpha_{\rm dS}&=H\left(\frac{3}{2}-\sqrt{\frac{9}{4}-\frac{\mu^2}{H^2}}\right),\\
&=\frac{\mu^2}{3H}+\mathcal{O}\!\left(\frac{\mu^4}{H^3}\right),
\qquad \mu\ll H.
\end{split}
\end{equation}
so the spectral gap can be made arbitrarily small as $\mu\to0$. With the present couplings one has $H\simeq 5.26\times10^{-2}$, so already at $\mu=0.01$ this gives $\alpha_{\rm dS}\simeq 6.36\times10^{-4}$. For a representative small three-horizon black hole with $(M,Q)=(1,1)$, the three positive roots are $(r_-,r_h,r_c)\simeq(0.758,1.481,17.784)$ and $\kappa_-\simeq 0.1269$, hence $\beta_{\rm SCC}\simeq \alpha_{\rm dS}/\kappa_-\approx 5.0\times10^{-3}\ll 1/2$ for this light field. In that sense the SCC bound is indeed automatically satisfied for sufficiently small black holes, precisely because a very slowly damped de Sitter-like branch controls the gap, in agreement with the general picture emphasized in Ref.~\cite{KonoplyaZhidenko:2022SCC}.

The same mechanism may persist beyond the strict small-black-hole limit, but here the evidence is suggestive rather than universal. For example, the three-horizon backgrounds $(M,Q)=(1.5,1.7)$ and $(2.0,3.2)$ have $(r_-,r_h,r_c)\simeq(0.942,2.907,17.062)$ and $(2.144,3.836,16.230)$, with $r_h/r_c\simeq 0.170$ and $0.236$, and with $\kappa_-\simeq 0.356$ and $0.146$, respectively. Even if one takes the larger nonzero de Sitter damping scale $\alpha_{\rm dS}\sim H$ associated with the leading nontrivial massless mode, one still gets $H/\kappa_-\simeq 0.15$ and $0.36$, both below the smooth-data threshold $1/2$ and certainly below the rough-data threshold $1$. This suggests that SCC remains satisfied in a nontrivial part of the three-horizon sector, not only in the tiniest-black-hole corner. On the other hand, the ratio is not monotonic across parameter space: near-extremal examples such as $(M,Q)=(2.0,3.5)$ have $\kappa_-\simeq 3.54\times10^{-2}$, so the naive estimate $H/\kappa_-\simeq 1.48$ already exceeds $1/2$. We therefore do not regard universal SCC protection for all three-horizon Proca black holes as established by the present data. Away from the small-black-hole regime, the dominant de Sitter-like frequencies should be computed explicitly in the charged three-horizon domain before making a global claim.

\section{Conclusions}

In this work we formulated and analyzed the problem of massive scalar perturbations of the full asymptotically de Sitter black-hole solution in generalized Proca theory. By keeping both the black-hole parameters $(M,Q)$ and the Proca-sector combinations $(\alpha,\beta,c_1,\lambda)$ explicit, we were able to track simultaneously the effects of primary hair, horizon structure, and the dynamically generated de Sitter scale. At the level of the effective potential, the comparison between the asymptotically flat and asymptotically de Sitter branches already reveals a qualitative difference: in the flat representative case the massive-field plateau eventually destroys the barrier, whereas in the de Sitter branch the potential vanishes at both horizons and retains a barrier throughout the static patch. The exact pure de Sitter spectrum therefore provides a natural analytic reference for identifying the de Sitter-like sector of the full black-hole problem.

The quasinormal-mode analysis across the five representative black-hole benchmarks shows a coherent pattern. At fixed background, increasing the scalar mass raises ${\rm Re}\,\omega$ and lowers $|{\rm Im}\,\omega|$ until a background-dependent damping plateau is reached. Increasing the black-hole size strongly softens the ringing, while increasing the Proca-hair parameter $Q$ at fixed $M$ slightly increases the oscillation frequency and decreases the damping. The angular-momentum splitting is appreciable only for light fields and becomes very small in the large-$\mu$ regime. For the reference neutral background, the first overtones follow the expected $(n+\frac{1}{2})$ pattern, and the time-domain Prony extraction agrees very well with the WKB data, confirming the consistency of the frequency-domain and time-domain descriptions.

These numerical trends admit a simple analytic explanation in the large-mass regime. When $\mu^2\gg \ell(\ell+1)/r^2$ and $\mu^2\gg |f'(r)/r|$, the peak of the effective potential is governed by the maximum of the metric function itself, and the quasinormal frequencies reduce to the compact asymptotic form $\omega_{n\ell}=\mu\,\Omega_R-i(n+\frac{1}{2})\Omega_I+\mathcal{O}(\mu^{-1})$. The leading damping rate is then independent of both $\mu$ and $\ell$, while the real part grows linearly with $\mu$. Direct comparison with the tabulated data shows that this approximation reproduces the large-mass spectrum with high accuracy. Physically, this means that, for the black-hole spectra analyzed here, the damping rate decreases strongly as the field mass grows but approaches a nonzero geometry-dependent constant rather than vanishing, so genuine quasi-resonances do not develop.

Our results also clarify the status of the strong-cosmic-censorship diagnostic. The five benchmark geometries used for the main WKB tables possess only an event horizon and a cosmological horizon, so the ratio $\beta_{\rm SCC}=\alpha_{\rm gap}/\kappa_-$ is not defined for those points directly. However, for the same Proca couplings there exists a charged three-horizon sector, and there the de Sitter-like branch provides the natural long-lived contribution to the spectral gap. For sufficiently small black holes this branch can be parametrically weakly damped, so that the SCC bound is automatically satisfied; the same tendency appears to persist in a nontrivial part of the wider charged sector, although near-extremal configurations indicate that it is not obviously universal. A definitive global statement about SCC in this model therefore requires an explicit computation of the dominant de Sitter-like modes throughout the full three-horizon region. Finally, the quasinormal modes obtained in the present paper can also be used to estimate the grey-body factors through the correspondence developed in Refs.~\cite{Konoplya:2024lir,Konoplya:2024vuj} and tested in a number of works~\cite{Bolokhov:2024otn,Malik:2024cgb,Lutfuoglu:2025mqa,Konoplya:2010vz,Dubinsky:2024vbn,Lutfuoglu:2025ldc,Malik:2025dxn,Bolokhov:2026eqf,Lutfuoglu:2025eik,Malik:2024wvs,Lutfuoglu:2025kqp}. More broadly, a systematic scan of $(M,Q,\alpha,\beta,c_1,\lambda,\mu,\ell)$ and a dedicated analysis of the charged three-horizon sector are the natural next steps.

\begin{acknowledgments}
The author would like to thank R. A. Konoplya for very helpful discussions.
\end{acknowledgments}

\bibliographystyle{apsrev4-1}
\bibliography{ProcaMassiveBH,referencesProca}
\end{document}